\documentclass[aip,jcp,reprint]{revtex4-1}

\usepackage[utf8]{inputenc}

\usepackage{amsmath}
\usepackage{amsfonts}
\usepackage{amssymb}
\DeclareMathOperator{\sgn}{sgn}

\usepackage{float}
\usepackage{graphicx}
\usepackage{colortbl}
\usepackage{color}

\begin{document}

\title{Angle-resolved effective potentials for disk-shaped molecules}

\author{Thomas Heinemann}
\email[]{thomas.heinemann@tu-berlin.de}
\affiliation{Institut für Theoretische Physik, Technische Universität Berlin, Hardenbergstr. 36, 10623 Berlin, Germany}
\author{Karol Palczynski}
\email[]{karol.palczynski@helmholtz-berlin.de}
\affiliation{Institut für Physik, Humboldt Universität zu Berlin, Newtonstraße 15, 12489 Berlin, Germany}
\affiliation{Helmholtz Zentrum Berlin (HZB), Institute of Soft Matter and Functional Materials, Hahn-Meitner Platz 1, 14109 Berlin, Germany}
\author{Joachim Dzubiella}
\email[]{joachim.dzubiella@helmholtz-berlin.de}
\affiliation{Institut für Physik, Humboldt Universität zu Berlin, Newtonstraße 15, 12489 Berlin, Germany}
\affiliation{Helmholtz Zentrum Berlin (HZB), Institute of Soft Matter and Functional Materials, Hahn-Meitner Platz 1, 14109 Berlin, Germany}
\author{Sabine H. L. Klapp}
\email[]{klapp@physik.tu-berlin.de}
\affiliation{Institut für Theoretische Physik, Technische Universität Berlin, Hardenbergstr. 36, 10623 Berlin, Germany}


\begin{abstract}
We present an approach for calculating coarse-grained angle-resolved effective pair potentials for uniaxial molecules. 
For integrating out the intramolecular degrees of freedom we apply umbrella sampling and steered dynamics techniques
in atomistically-resolved molecular dynamics (MD) computer simulations.
Throughout this study we focus on disk-like molecules such as coronene. To develop the methods we focus 
on integrating out the van-der-Waals and intramolecular interactions, while 
electrostatic charge contributions are neglected.
The resulting coarse-grained pair potential reveals a strong temperature and angle dependence. In the next step we fit the numerical data with various Gay-Berne-like potentials to be used in more efficient simulations on larger scales.
The quality of the resulting coarse-grained results is evaluated by comparing their pair and many-body structure as well as some thermodynamic quantities self-consistently
to the outcome of atomistic MD simulations of many-particle systems. We find that angle-resolved potentials are essential not only to
accurately describe crystal structures but also for fluid systems where simple isotropic potentials start to fail already for low to
moderate packing fractions.
Further, in describing these states it is crucial to take into account the pronounced temperature dependence arising in 
selected pair configurations due to bending fluctuations.
\end{abstract}

\maketitle

\section{Introduction}

In the past decades much effort has been devoted to define effective Hamiltonians~\cite{Likos2001,Hansen2002} for many-particle systems such as, e.g.,
systems of water clusters~\cite{Wu2010}, dissolved ions~\cite{Kalcher2010}, polymers~\cite{Baschnagel1991}, phospholipids~\cite{Izvekov2005} and 
(bio-)molecules like protein-DNA complexes~\cite{Villa2004}.
Typically, these effective Hamiltonians are restricted to pair terms, where the effective pair potentials 
are either suggested heuristically~\cite{Groot1997} or derived by a systematic coarse-graining procedure, implying that ``irrelevant'' degrees of freedom
are integrated out. One main motivation behind the construction 
of such effective potentials is to enable computer simulations on length and time scales larger than those accessible for the underlying original system.
This is achieved, on the one hand, by considering fewer degrees of freedom, and on the other one hand, by the enhanced softness
of effective interactions \cite{Klapp2004}, which allows for larger time steps.

In most studies so far, the effective potentials are purely distance-dependent, where the distance considered is typically that between center of masses \cite{Young1997}. 
Interactions between {\em non-spherical} molecules are then described, e.g., by representing the
molecule as interconnected spherical beads~\cite{Izvekov2005,Ghosh2012}, an approach which seems particularly suitable for large, flexible molecules
such as polymers. 

In the present study, we consider effective interactions between anisotropic molecules which have, however, a well-defined shape and are characterized by 
uniaxial symmetry. For such systems we present 
a coarse-graining approach yielding effective pair potentials depending on both, distance and angular variables. Our overall aim is not only to provide
a recipe to calculate such a potential, but also to evaluate the importance of angular resolution of the potential relative to a simplified center of mass description.
We also aim to explore the dependence of the angle-resolved potential on temperature.

\begin{figure}
 \begin{center}
  \includegraphics[width=0.5\columnwidth]{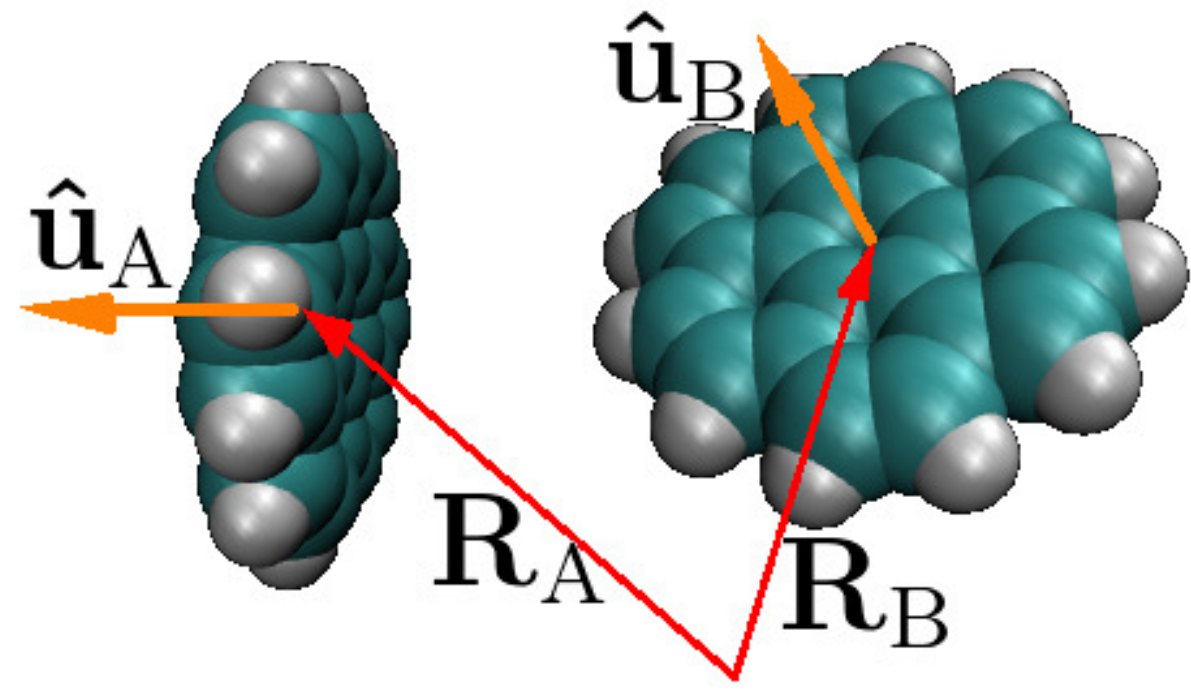}
 \end{center}
 \caption{\label{fig:twocoros}An exemplary configuration of two coronene molecules illustrating their internal atomic structure. Green and gray spheres
 refer to carbon and hydrogen atoms, respectively. The arrows stand for coarse-grained (vectorial) variables introduced in Sec.~\ref{sec:Reaction coordinates}. }
\end{figure}

As a candidate system we consider a pair of two disk-like molecules such as coronene. A sketch of the system is given in Fig.~\ref{fig:twocoros}.
Coronene is a conjugated organic molecule with a disk-like shape, for which the assumption of uniaxiality is
well justified \cite{Obolensky2007}. Moreover, coronene molecules have already been discussed 
as possible candidates for active layer compounds in photovoltaic applications~\cite{Blumstengel2008}. Indeed,
in organic solar cells, diskotic organic molecules such as triphenylenes, hexabenzocoronenes and their derivatives are quite common \cite{Andrienko2006}.
A prerequisite for advancing the functionality of organic solar cells is to understand the many-particle structures of the molecules involved. 
For example, for disk-like molecules such as coronene one expects the formation of columnar structures \cite{Chandrasekhar1993,Bushby2002},
indicating that any coarse-grained description of such systems {\it must} take into account the molecule's anisotropic shape.

The question is then which additional microscopic effects 
(beyond the anisotropic shape) need to be taken into account and how this should be done.
Indeed, on an atomistic level, coronene-coronene interactions are characterized by 
attractive van-der-Waals interactions, intramolecular flexibility
as well as Coulombic interactions stemming from the partial charges. Various coarse-grained models for coronene already exist; however, none of these
includes all presumably important features. For example, 
von Lilienfeld and Andrienko~\cite{Lilienfeld2006} have suggested a 
coronene pair potential which is based on quantum chemical calculations. However, this potential takes into account the face-to-face configuration alone.
Thus, the potential does not take into account the full configurational space.
Babadi et al.~\cite{Babadi2006} have proposed a pair potential which corresponds to fit 
according to an ellipsoidal soft potential suggested in Ref.~\onlinecite{Everaers2003}. This potential is indeed
angle-dependent but does not depend on temperature and, thus, neglects entropic effects.
In yet another study, Obolensky et al~\cite{Obolensky2007} proposed a uniaxial model, where each coronene molecule
is represented as a collection of charged rings. Thus, the model takes into account the electrostatic contributions to the effective potential. One drawback, however, is that
the evaluation of the resulting potential is numerically quite involved and therefore inconvenient for many-particle simulations.
Indeed, the calculations in Ref.~\onlinecite{Obolensky2007} rather focus on dimer configurations. Moreover, 
this specific model does not include the impact of temperature.

The above examples show that
finding the "right" coarse-grained coronene-coronene interactions is not straightforward.
In the present study, we simplify the task and concentrate on deriving effective potentials stemming from non-electrostatic interactions alone.
This restriction implies
that we cannot describe the realistic crystal con\-fi\-gu\-ra\-tion of coronene, which corresponds to a herringbone structure \cite{Robertson1944}. We note that a correct 
description of the electrostatics would 
include not only dealing with long-range Coulombic potentials,
but also treating polarization effects, that is, 
differences in the molecular charge distributions within dilute systems (i.e., isolated molecules), on the one hand, 
and dense systems, on the other hand \cite{Fedorov2012}. Here we avoid this task and focus on
the remaining challenges, that is, the description of angle dependency and temperature dependency due to the non-electrostatic interactions. Our goal is to provide a versatile "recipe"
which may be applied to a class of anisotropic, uniaxial molecules. The coronene molecule is used as an {\em example} for establishing our approach. 

From a methodological point of view, we employ the ``classical'' statistical-mechanical route first suggested by
Kirkwood~\cite{Kirkwood1935}, who introduced the potential of
mean force (PMF).
The PMF is defined as the difference of the free energy profiles between two molecular configurations. These free energy profiles
can be calculated by performing a Boltzmann inversion of the corresponding probability distribution functions gained in corresponding atomistic simulations.
For a system composed of only two molecules the free energy profiles then lead directly 
to effective pair potentials, which include entropic contributions. Therefore, the
resulting potential depends on the temperature.

Other routes suggested in the literature are based on force-matching (see, e.g., Ref.~\onlinecite{Loewen1993} for charged particles in solvent), a method which can be 
extended towards internal degrees of freedom \cite{Ercolessi1994} and to multiscale systems \cite{Izvekov2004,Izvekov2005}. Further, 
fundamentally different approaches are the reverse Monte Carlo~\cite{McGreevy1988} technique, the iterative Boltzmann inversion scheme~\cite{Tschoep1998,Tschoep1998b,Silbermann2006}
or integral equation schemes~\cite{Li2005} based on structural properties.
The present study corresponds to a generalization of the original Kirkwood route towards a two-particle system with spatial and angular degrees of freedom.

As a method to generate the underlying probability distribution functions we use all-atom Langevin dynamics, i.e., Molecular Dynamics coupled to a heat bath.
To overcome sampling problems we use and compare two different methods, each having its own advantages. The first one
is the umbrella sampling method \cite{Torrie1974,Torrie1977} involving static bias potentials, combined with 
the weighted histogram analysis method (WHAM) \cite{Kumar1992,Roux1995}. 
The second method is referred to as steered dynamics \cite{Mulders1996,Izrailev1999}, which is inspired from experiments 
where large molecules are stretched and then the rupture force is measured \cite{Grubmueller1996}.
This method has already been used, e.g., in ligand-receptor simulations~\cite{Izrailev1999}.

For both sampling methods, we parametrize the resulting potential curves in terms of a modified Gay-Berne potential. This step facilitates 
simulations of large ensembles at different packing fractions and temperatures. By comparing 
the resulting thermodynamic quantities and phase behavior with that of the underlying all-atom system, we can evaluate the quality of the coarse-grained potentials.
We find that the angle-dependence of the potential is important not only in dense, liquid-crystalline states, but already
at intermediate densities.

The remainder of this article is organized as follows. Section~\ref{sec:Effective interaction of coronene molecules} is devoted to our methods, including the definition of coarse-grained variables (Sec.~\ref{sec:Reaction coordinates}), the definition of the effective pair potential
via partition sums (Sec.~\ref{sec:Definition of the effective pair potential}), and a description of the sampling methods (Sec.~\ref{sec:Sampling methods}).
In Sec.~\ref{sec:Results for the effective potential} we present the numerical results for effective potentials in different angular configurations and at different temperatures. The fit of the numerical potentials in terms of a Gay-Berne potential is discussed in Sec.~\ref{sec:Parametrization in terms of a Gay-Berne potential}. In Sec.~\ref{sec:Many particle simulations} we discuss the results from many-particle simulations based on the effective potentials and all-atom simulations, focusing on the phase behavior of the system. Finally, conclusions are given in Sec.~\ref{sec:Conclusion}.

\section{Effective interaction of coronene molecules}\label{sec:Effective interaction of coronene molecules}
\subsection{Atomic system}\label{sec:Atomic system}

The system of interest consists of two atomistically detailed coronene molecules (${\text{C}_{24}\text{H}_{12}}$) in a large cubic box
${V=l^3}$ with periodic boundary conditions ($l$ is the boxlength).
Each coronene molecule contains ${N=36}$ atoms.
Every atom $i$ in our model system is represented by a point mass $m_i$ at the position $\mathbf{r}_i$.
The interactions between all atoms are described by Lennard-Jones (LJ) potentials for non-bonded interactions and
harmonic potentials for the intramolecular bond-, angular- and dihedral-interactions.
For the present study all Coulomb interactions are set to zero.
The potential energy as a function of all ${2\,N}$ atomic coordinates can then be written as  
\begin{eqnarray}\label{eqn:potentialenergy} 
  U\left( \{\mathbf{r}_i\}_{i=1,\dots,2N}   \right) & = & \sum_{i,j}\left[\left(\frac{(C_{6})_{ij}}{r_{ij}}\right)^{6}-\left(\frac{(C_{12})_{ij}}{r_{ij}}\right)^{12}\right]\nonumber\\
 &  & +\frac{1}{2} \sum_{i,j} K_{ij}^{b}\left(r_{ij}-r_{\mathrm{\text{eq}}}\right)^{2}\\
 &  & +\frac{1}{2} \sum_{i,j,k} K_{ijk}^{\theta}\left(\theta_{ijk}-\theta_{\mathrm{\text{eq}}}\right)^{2}\nonumber\\
 &  & +\frac{1}{2} \sum_{i,j,k,l} K_{ijkl}^{\phi}\left[1+\cos\left(n\phi_{ijkl}-\epsilon \right)\right],\nonumber
\end{eqnarray} 
where $(C_{6})_{ij}$ and $(C_{12})_{ij}$ are the LJ parameters between atoms $i$ and $j$.
Atomic distances are denoted with ${r_{ij}=\left| \mathbf{r}_i-\mathbf{r}_j\right|}$.
Further, $K_{ij}^{b}$ and $K_{ijk}^{\theta}$ are force constants
for the intramolecular bond- and angular- interactions,  and $r_{\mathrm{\text{eq}}}$ and $\theta_{\mathrm{\text{eq}}}$ are the corresponding equilibrium bond lengths and bond angles, respectively.
The quantity $K_{ijkl}^{\phi}$ is a dihedral parameter and $\phi_{ijkl}$ is the corresponding dihedral angle, while $\epsilon$ serves as a phase angle which is either
$0^\circ$ or $180^\circ$.
The factor $n$ appearing in the last term stands for the proper-dihedral multiplicity~\cite{Wang2004}.
All  parameter values are taken from the generalized Amber force field, designed for organic molecules~\cite{Wang2004}.
The positions of the atoms evolve in time according to all-atom Langevin dynamics, that is,
\begin{align}\label{eqn:langevindynamics}
 m_i\ddot{\mathbf{r}}_i(t)=-\nabla_i U(\{\mathbf{r}\},t)  -\gamma \,m_i \dot{\mathbf{r}}_i(t)+\sqrt{2\gamma \, k_{\text{B}} T\,m_i} \mathbf{X}(t)\text{.}
\end{align}
In Eq.~\eqref{eqn:langevindynamics}, ${\mathbf{X}(t)}$ is a vector whose components ${X_{\alpha}(t)}$ are Gaussian random numbers with ${\left<X_{\alpha}(t)\right>=0}$,
${\left<X_{\alpha}(t)\, X_{\beta}(t')\right>= \delta_{\alpha\beta} \, \delta (t-t')}$.
The friction constant is denoted with $\gamma$ which is set to ${0.25\,\mathrm{ps}^{-1}}$, and the two non-conservative forces are coupled via the fluctuation-dissipation theorem.
In the actual numerical simulations the equations of motion [see Eq.~\eqref{eqn:langevindynamics}] are supplemented by constraints or bias potentials as described in Sec.~\ref{sec:Sampling methods}.
The resulting set of equations is solved with the GROMACS~\cite{VanDerSpoel2005} simulation package, using version 4.5.4 for the steered dynamics and version 4.5.5~\cite{bug} for the umbrella sampling.
The cutoff-lengths for the atomic LJ interactions are set to ${2\,\mathrm{nm}}$. The simulations last for ${275\,\mathrm{ns}}$ with an integration time step of ${1\,\mathrm{fs}}$.
To calculate histograms trajectories are extracted every ${10\,\mathrm{fs}}$.

\subsection{Reaction coordinates}\label{sec:Reaction coordinates}

An important step in any coarse-graining procedure is to define variables that represent the coarse-grained, mesoscopic system.
Here we describe each coronene molecule by the center of mass position $\mathbf{R}$ and an orientation vector  $\mathbf{\hat{u}}$,
pointing along the axis related to the largest eigenvalue of the atomistic tensor of moments of inertia.
A configuration of the coronene ``dimer'' consisting of the two individual molecules A and B is therefore defined by the four three-dimensional vectors $\mathbf{R}_{\text{A}}$, $\mathbf{R}_{\text{B}}$, $\mathbf{\hat{u}}_{\text{A}}$,  $\mathbf{\hat{u}}_{\text{B}}$ (see Fig.~\ref{fig:twocoros}).
This choice of coarse-grained variables seems most natural
due to several reasons: first, the inertia tensor is symmetric implying that the resulting coarse-grained variables do not change under simultaneous change
of the atomistic masses. Second, the chosen set of variables is compatible with the variables used in the Gay-Berne model, which we will later use to parametrize
our coarse-grained potential (see Sec.~\ref{sec:Parametrization in terms of a Gay-Berne potential}). Third, the center-of-mass description provides a particularly comfortable route to calculate the virial pressure.

The number of variables describing the coronene ``dimer'' can be further reduced by transforming to the body-fixed frame and  using the head-tail symmetry of the particles.
Moreover, we require the effective interaction of the two molecules to have chiral symmetry (i.e., it should be invariant against mirroring the dimer system).
This finally leads to a set of four reaction coordinates: 
\begin{align}\label{eqn:reaction coordinates}
R&=\left|\mathbf{R}_{\text{B}}-\mathbf{R}_{\text{A}}\right| \text{,}\nonumber\\
a&=\left|\mathbf{\hat{u}}_{\text{A}}\cdot\mathbf{\hat{R}}\right| \, \text{with} \, \mathbf{\hat{R}}= \frac{\mathbf{R}_{\text{B}}-\mathbf{R}_{\text{A}}}{\left|\mathbf{R}_{\text{B}}-\mathbf{R}_{\text{A}}\right|}\text{,}\nonumber\\
b&=\left|\mathbf{\hat{u}}_{\text{B}}\cdot\mathbf{\hat{R}}\right| \text{,} \nonumber\\
c&=\sgn(\mathbf{\hat{u}}_{\text{A}}\cdot\mathbf{\hat{R}})\,\sgn(\mathbf{\hat{u}}_{\text{B}}\cdot\mathbf{\hat{R}})\,\mathbf{\hat{u}}_{\text{A}}\cdot\mathbf{\hat{u}}_{\text{B}}\text{.}
\end{align}
The coordinate $R$ stands for the molecular distance, while $a$, $b$ and $c$ represent angular configurations of the dimer.
In the last line of Eq.~\eqref{eqn:reaction coordinates}, ``$\sgn$'' is the sign function defined as ${\sgn(x)= 1}$ for ${x>0}$; ${\sgn(x)= -1}$ for ${x<0}$; and ${\sgn(0)= 0}$.
For further investigation we also introduce the corresponding functions that map the atomic description directly on the coarse-grained description.
They are denoted with $\tilde{R}$, $\tilde{a}$, $\tilde{b}$ and $\tilde{c}$.
\subsection{Definition of the effective pair potential}\label{sec:Definition of the effective pair potential}

In this section we derive an effective pair potential, which depends on the reaction coordinates defined in Eq.~\eqref{eqn:reaction coordinates}.
The coronene dimer, which consists of $2N$ atoms (with the atoms $\mathbf{r}_1$, \dots, $\mathbf{r}_N$ belonging to molecule A and the rest belonging to molecule B), leads
to the following canonical configuration integral
\begin{align}\label{eqn:Zc}
 Z_{\text{c}}= \frac{1}{\alpha^{6N}} \int_{V} d\mathbf{r}_1\dots\int_{V} d\mathbf{r}_{2N}  \, e^{-\beta\, U(\{\mathbf{r}\})}\text{,}
\end{align}
where $\alpha$ has the dimension of length.
Each atomic con\-fi\-gu\-ra\-tion ${\{\mathbf{r}\}=\{\mathbf{r}_i\}_{i=1,\dots,2N}}$ corresponds to a unique set of reaction coordinates, i.e.
\\
${1=\int_{0}^l \!  dR \, \delta(R\!-\!\tilde{R}(\{\mathbf{r}\}))}$, \dots,${1 =\int_{\mathbb{R}}  \! dc \, \delta\left(c\!-\!\tilde{c}(\{\mathbf{r}\})\right)}$.
Therefore, the canonical configuration integral can be written as an integration over the reaction coordinates $R$, $a$, $b$ and $c$, yielding
\begin{multline}\label{eqn:Zc2}
Z_{\text{c}}=\frac{1}{\alpha}\int_{0}^l \! \! \! \! dR\dots\! \int_{\mathbb{R}} \! \! dc  \,\frac{1}{\alpha^{6N-1}}   \int_{V} \!\! d\mathbf{r}_1\dots\! \!\int_{V} \! d\mathbf{r}_{2N}  \\ \delta(R\!-\!\tilde{R}(\{\mathbf{r}\}))\dots\delta\left(c\!-\!\tilde{c}(\{\mathbf{r}\})\right)\, e^{-\beta\, U(\{\mathbf{r}\})}\text{.}
\end{multline}
In Eq.~\eqref{eqn:Zc2} the appearance of ${\int_{V} \!\! d\mathbf{r}_1\dots\! \!\int_{V} \! d\mathbf{r}_{2N}  \,\delta(R\!-\!\tilde{R}(\{\mathbf{r}\}))\dots\delta\left(c\!-\!\tilde{c}(\{\mathbf{r}\})\right)}$
indicates the constrained integration over the subclass of microstates, which corresponds to the reaction coordinates $R$, $a$, $b$ and $c$.
We next introduce the configuration integral for a fixed mesoscopic configuration
\begin{multline}\label{eqn:ZRabc}
{\mathcal Z}_{\text{c}}(R,a,b,c)= \frac{1}{\alpha^{6N-1}}\int_{V} \!\! d\mathbf{r}_1\dots\! \!\int_{V} \! d\mathbf{r}_{2N} \\ \delta(R\!-\!\tilde{R}(\{\mathbf{r}\}))\dots\delta\left(c\!-\!\tilde{c}(\{\mathbf{r}\})\right)\, e^{-\beta\, U(\{\mathbf{r}\})}\text{.}
\end{multline}
Combining Eqs.~\eqref{eqn:Zc2} and~\eqref{eqn:ZRabc} we find
\begin{align}\label{eqn:Zsmall}
 Z_{\text{c}}=\frac{1}{\alpha}\int_{0}^l \! \! \! \! dR\dots\! \int_{\mathbb{R}} \! \! dc  \,  {\mathcal Z}_{\text{c}}(R,a,b,c)\text{.}
\end{align}
At this point, it seems plausible to define an effective interaction potential (or rather, a distance dependent free energy profile) simply by
taking the logarithm of ${{\mathcal Z}_{\text{c}}(R,a,b,c)}$. However, closer inspection of the definition~\eqref{eqn:ZRabc} reveals that ${{\mathcal Z}_{\text{c}}(R,a,b,c)}$ still depends on
the values of $R$, $a$, $b$, $c$ even if the distances considered are much larger than the range of the all-atom potential ${U(\{\mathbf{r}\})}$
[see Eq.~\eqref{eqn:potentialenergy}].
This is clearly unphysical. The reason for the problem is that different values of  $R$, $a$, $b$, $c$ imply different numbers of sampled microstates.
We therefore introduce a new effective configuration integral ${{\mathcal Z}_{\text{eff}}(R,a,b,c)}$, which is normalized by the amount of configuration space volume, that is,
\begin{align}\label{eqn:Zeff}
 {\mathcal Z}_{\text{eff}}(R,a,b,c)=\frac{{\mathcal Z}_{\text{c}}(R,a,b,c)}{{\mathcal Z}_{\text{mic}}(R,a,b,c)}\text{,}
\end{align}
where ${\mathcal Z}_{\text{mic}}$ is a ``microcanonical'' integral, which counts the number of microstates belonging to a fixed set of coarse-grained variables
$R$, $a$, $b$, $c$.
Specifically,
\begin{multline}\label{eqn:Zmic}
{\mathcal Z}_{\text{mic}}(R,a,b,c)=\frac{1}{\alpha^{6N-1}}\int_{V} \!\! d\mathbf{r}_1\dots\! \!\int_{V} \! d\mathbf{r}_{2N} 
\\
\delta(R\!-\!\tilde{R}(\{\mathbf{r}\}))\dots\delta\left(c\!-\!\tilde{c}(\{\mathbf{r}\})\right)\text{.}
\end{multline}
This expression can be further simplified (see Appendix~\ref{sec:Configurational entropy for (R,a,b,c)}).
The resulting effective configuration integral is independent of its variables at large distances,
that is,
\begin{align}\label{eqn:Zeff2}
 {\mathcal Z}_{\text{eff}}(R,a,b,c)\xrightarrow{R\rightarrow \infty} \Lambda= \text{const}\text{.}
\end{align}
Based on ${\mathcal Z}_{\text{eff}}$, we now define an effective potential via
\begin{align}\label{eqn:Ueff}
U_{\text{eff}}(R,a,b,c)=-\frac{1}{\beta} \, \ln\left[\frac{{\mathcal Z}_{\text{eff}}(R,a,b,c)}{\Lambda}\right]\text{.}
\end{align}
From Eqs.~\eqref{eqn:Zeff2} and~\eqref{eqn:Ueff}, it follows that ${U_{\text{eff}}(R,a,b,c) \rightarrow 0}$ for ${R \rightarrow \infty}$, as one would expect.
We determine the quantity ${{\mathcal Z}_{\text{c}}(R,a,b,c)}$ entering ${\mathcal Z}_{\text{eff}}$ [see Eq.~\eqref{eqn:Zeff}] by solving Eq.~\eqref{eqn:ZRabc} through numerical sampling methods as described in Section~\ref{sec:Sampling methods}.
The denominator in Eq.~\eqref{eqn:Zeff}, ${{\mathcal Z}_{\text{mic}}(R,a,b,c)}$, can be associated to a Boltzmann entropy
\begin{align}\label{eqn:confentropy}
S_{\text{B}}(R,a,b,c)=k\, \ln \left[\Lambda\, {\mathcal Z}_{\text{mic}}(R,a,b,c)\right]\text{.} 
\end{align}
Taken altogether, we can thus interpret $U_{\text{eff}}$ as a configuration-dependent free energy, corrected by the Boltzmann entropy related to the configuration space
spanned by the coarse-grained variables.
If the distance $R$ is the only variable on the mesoscopic level, the entropy reduces to 
${S_{\text{B}}(R)=2\,k\, \ln(R/\alpha) + \text{const}}$, as outlined in Ref.~\onlinecite{Neumann1980}.
\subsection{Sampling methods}\label{sec:Sampling methods}
To calculate the effective potential [see Eq.~\eqref{eqn:Ueff}] from the trajectories of the atomic system we introduce the histogram function, 
\begin{align}\label{eqn:P(R,a,b,c)}
 P(R,a,b,c)&=\frac{{\mathcal Z}_{\text{c}}(R,a,b,c)}{\alpha\,Z_{\text{c}}}\\
 &=\left< \delta(R-\tilde{R})\,\delta(a-\tilde{a})\,\delta(b-\tilde{b})\,\delta(c-\tilde{c})\right>\text{,}\nonumber
\end{align}
where the brackets $<\dots>$ in the second line denote an ensemble average in the atomic system and we have used the definitions Eqs.~\eqref{eqn:Zc} and ~\eqref{eqn:Zsmall}.
With Eq.~\eqref{eqn:P(R,a,b,c)}, the effective pair potential can be written as
\begin{align}\label{eqn:ueffpractical}
 U_{\text{eff}}(R,a,b,c)=-\frac{1}{\beta}\ln\left[ \frac{P(R,a,b,c)}{{\mathcal Z}_{\text{mic}}(R,a,b,c)}\, \frac{\alpha\,Z_{\text{c}}}{\Lambda}  \right] \text{.}
\end{align}
To perform the configurational sampling, i.e. to actually calculate the function $P(R,a,b,c)$, we perform all-atom Langevin dynamics simulations as described in Sec.~\ref{sec:Atomic system}.
The noise term in the corresponding equations of motion [see Eq.~\eqref{eqn:langevindynamics}] generates internal translational and rotational motion, i.e., translational motion
of individual atoms and rotations of the entire molecule around the molecules’ center of mass.
However, standard sampling is hampered by the fact that the two molecules strongly attract each other.
In the next paragraphs we describe two methods to overcome this drawback by restraining or constraining the molecules to a certain distance $R$, while the orientational motion is undisturbed.
\subsubsection{Umbrella sampling}\label{sec:Umbrella sampling}
In the framework of umbrella sampling~\cite{Torrie1974,Torrie1977}, the Hamiltonian of the system is supplemented by a bias potential to support the sampling in different regions of con\-fi\-gu\-ra\-tion space.
Together with the weighted histogram analysis method~\cite{Kumar1992} (WHAM) umbrella sampling was already used to construct purely distance-dependent effective molecular pair potentials, e.g. for methane in aqueous solution~\cite{Young1997}.
In this article we use bias potentials that correspond to harmonic springs. Specifically,
\begin{align}\label{eqn:V_k}
V_k(\{\mathbf{r}\})=\frac{c_k^{\text{spring}}}{2} \left(\tilde{R}(\{\mathbf{r}\})-R_k^{\text{eq}}\right)^2 
\end{align}
for $R_k^{\text{eq}}>R_{k-1}^{\text{eq}}, k=1,\dots,N_w$.
Each of the $N_w$ springs acts on the molecular centers of mass and is used for one specific simulation, called umbrella window run.
The equilibrium length for each spring, $R_k^{\text{eq}}$, and the spring constants, $c_k^{\text{spring}}$, are chosen to guarantee a strong overlap of
the $R$-dependent biased histogram functions $P_k^{\text{bias}}(R)=\left< \delta(R-\tilde{R}) \right>_{k}$
for neighboring umbrella windows (k, k+1).
The brackets ${\left< \dots \right>_{k}}$ denote an ensemble or time average in the umbrella window $k$.
Out of $P_k^{\text{bias}}(R)$, we can obtain purely distance-dependent, unbiased histograms $P(R)$ by using the one-dimensional WHAM equations given in Appendix~\ref{sec:The WHAM equations}.
In principle, it is possible to extend these equations to the multidimensional case~\cite{Kumar1992} involving additional reaction coordinates $a$, $b$, $c$.
In our case the bias potential is a function of $R$ alone [see Eq.~\eqref{eqn:V_k}] while we are interested in the four-dimensional histogram ${P(R,a,b,c)}$.
Therefore we employ Eqs.~\eqref{eqn:WHAMoriginalb} and~\eqref{eqn:WHAMoriginalc} together with the following decomposition of the full unbiased histogram function
\begin{align}\label{eqn:unbiasedhistogramdecompostionRabc}
P(R,a,b,c)=\sum_{k=1}^{N_w} \gamma_k(R) \cdot P_k^{\text{bias}}(R,a,b,c)\text{,}
\end{align}
where the $\gamma_k$ represent $R$-dependent coefficients defined in Eq.~\eqref{eqn:WHAMoriginalc}.
A similar strategy has been recently used in Ref.~\onlinecite{Ghosh2012} where the goal was to obtain the effective pair potential of a methanol pair dissolved in water with two reaction coordinates.
\vspace*{1em}
\subsubsection{Steered dynamics}\label{sec:Steered dynamics}
In a steered dynamics simulation~\cite{Mulders1996,Izrailev1999} a reaction coordinate is changed in time by applying external constraining forces.
Here we pull one molecule away from the other one along the connecting vector $\mathbf{R}$.
Specifically, the molecular distance is steered according to the following law
\begin{align}\label{eqn:R(t)}
 R(t)=R_0+c_{\text{pull}} \cdot t\text{,}
\end{align}
where $c_{\text{pull}}$ is the pull rate. The latter is so small, that the entire simulation can be seen as a quasi-static process.
Therefore, given a small time interval, the system evolves according to the so-called constrained-reaction-coordinate-dynamics ensemble~\cite{Carter1989}.
Meanwhile all rotational degrees of freedom remain unconstrained.
The scheme in Eq.~\eqref{eqn:R(t)} implies an equal weighting of all values of $R$. Therefore the corresponding $R$-dependent histogram function ${P(R)}$ forms a flat distribution.
That means that ${P(R)}$ does not provide any information about the distance-dependent effective potential.
Hence ${P(R,a,b,c)}$ is not measureable in that way.
In order to overcome this drawback we factorize the unconstrained histogram function, as follows
\begin{align}\label{eqn:pdecomp}
 P(R,a,b,c)=P(R) \cdot  \left(\frac{P(R,a,b,c)}{P(R)}\right)\text{.}
\end{align}
In Eq.~\eqref{eqn:pdecomp} the distribution ${P(R)}$ can be calculated via the free energy ${A(R)}$. Specifically, one has
\begin{align}\label{eqn:P(R)}
 P(R)=\left< \delta(R-\tilde{R}) \right>=\frac{{\mathcal Z}_{\text{c}}(R)}{\alpha \, Z_{\text{c}}}=\frac{e^{-\beta A(R)}}{\alpha\,Z_{\text{c}}}\text{,}
\end{align}
where ${A(R)=-1/\beta \ln {\mathcal Z}_{\text{c}}(R)}$ can be determined through a thermodynamic integration~\cite{Kirkwood1935}.
\begin{widetext}
To calculate the remaining function ${P(R,a,b,c)}$ we use Eq.~\eqref{eqn:P(R,a,b,c)}, yielding
\begin{align}
 \frac{P(R,a,b,c)}{P(R)}&=\frac{{\mathcal Z}_{\text{c}}(R,a,b,c)}{{\mathcal Z}_{\text{c}}(R)}=\nonumber\\
 &=\frac{ \frac{1}{\alpha^{6N-1}}\int_{V} d\mathbf{r}_1\dots\int_{V} d\mathbf{r}_{2N} \, e^{-\beta\, U(\mathbf{r}_1,\dots,\mathbf{r}_{2N})}  \,  \delta(R-\tilde{R})      \,\delta(a-\tilde{a})\,\,\delta(b-\tilde{b})\,\,\delta(c-\tilde{c})   }      
 {  \frac{1}{\alpha^{6N-1}} \int_{V} d\mathbf{r}_1\dots\int_{V} d\mathbf{r}_{2N}\, e^{-\beta\, U(\mathbf{r}_1,\dots,\mathbf{r}_{2N})}  \,\delta(R-\tilde{R})   }\text{.}
\end{align}
\end{widetext}
The right hand side can be considered as an ensemble average over all atomic configurations, that correspond to the specific center of mass distance $R$.
Specifically,
\begin{align}\label{eqn:P(R,a,b,c|R)}
\frac{P(R,a,b,c)}{P(R)}&=\left<\delta(a-\tilde{a})\,\,\delta(b-\tilde{b})\,\,\delta(c-\tilde{c})\right>_{\text{ensemble}}^{\text{distance}=R}\\
&=P(R,a,b,c|R)\text{.}\nonumber
\end{align}
We can conclude that ${P(R,a,b,c|R)}$ is the conditional histogram function of the angle dependent reaction coordinates for a fixed distance $R$.
For small time intervals ${\Delta t}$ (small enough to ensure that $R$ is fixed, but large enough to sample the entire angular configuration space) we can 
replace the ensemble average in Eq.~\eqref{eqn:P(R,a,b,c|R)} by a time average, that is
\begin{align}\label{eqn:P(R,a,b,c|R)2}
 P(R,a,b,c|R) \rightarrow \left<\delta(a-\tilde{a})\,\,\delta(b-\tilde{b})\,\,\delta(c-\tilde{c})\right>^{\text{distance}=R}_{\Delta t} \text{.}
\end{align}
The smaller the pull rate $c_{\text{pull}}$, the larger ${\Delta t}$ can be.
Finally, by combining Eqs.~\eqref{eqn:pdecomp},~\eqref{eqn:P(R)} and~\eqref{eqn:P(R,a,b,c|R)2} we obtain the unconstrained function ${P(R,a,b,c)}$.
\subsubsection{Numerical details}\label{sec:Numerical details}

In order to determine the histogram functions ${P_k^{\text{bias}}(R)}$ and ${P_k^{\text{bias}}(R,a,b,c)}$ for umbrella sampling or ${P(R,a,b,c|R)}$ for steered dynamics, respectively,
we use the following scheme.
For both sampling methods the molecular distance $R$, is sub-divided into $800$ bins covering the interval ${[0.23\,\mathrm{nm}, 4.5\,\mathrm{nm}]}$.
To capture the angle dependence, we focus on eight configurations, 
namely the face-face, parallel weakly displaced, parallel displaced, T, herringbone, V, edge-edge and the cross configuration.
Graphical representations as well as explicit definitions in terms of the reaction coordinates $a$, $b$ and $c$ are given in Table~\ref{tab:tolerances} in Appendix~\ref{sec:Dimer configurations}.
In order to assign a set of coordinates ($a$, $b$, $c$) to one of these configurations, we use the tolerances given in this table.
To obtain a smooth result for the effective interactions ${U_{\text{eff}}(R,a,b,c)}$ for the previously introduced configurations, we reduce the number of bins 
in the angle-resolved unbiased histogram function ${P(R,a,b,c)}$ that determines ${U_{\text{eff}}(R,a,b,c)}$ [see  Eq.~\eqref{eqn:ueffpractical}]
to $100$.
This is realized by associating bin $x$ to bin ${y(x)=30.167\,\ln[x/30.167+1]}$, yielding ${y(0)=0}$, ${y(800)\approx 100}$ and ${y'(0)=1}$.
As a consequence the first bins are small, while bins for larger distances are bigger to account for the weaker sampling in these regions.
In the umbrella sampling simulations we use 50 umbrella windows ($k=0,\dots,49$) each longing for ${5.5\, ns}$.
The initial configuration is always set to the face-face configuration.
Each spring is characterized by a spring constant of ${c_k^{\text{spring}}=500 \, \mathrm{kJ}/\mathrm{mol}/ \mathrm{nm}}$ and an equilibrium length of ${R_k^{\text{eq}}=(0.253+k\cdot 0.055) \, \mathrm{nm}}$.
The WHAM-equations for the distance $R$ (see Appendix~\ref{sec:The WHAM equations}) are repeatedly solved until the change in the free energy constants $F_k$ is below ${10^{-5} \, \mathrm{kJ}/\mathrm{mol}}$.
Resulting weights ${\gamma(R)}$ serve as new weights in the multidimensional decomposition [see Eq.~\eqref{eqn:unbiasedhistogramdecompostionRabc}].
\par
The start distance in the steered dynamics simulations is set to ${0.253 \, \mathrm{nm}}$ in a face-face constellation.
Then, the second molecule is pulled away from the first with a rate of ${c_{\text{pull}} = 10^{-5} \, \mathrm{nm}/\mathrm{ps}}$.

\section{Results for the effective potential}\label{sec:Results for the effective potential}
In this section we apply the two sampling methods introduced before to calculate the effective potential of a coronene dimer system at various temperatures $T$.
We first consider the angle-averaged potential ${U_{\text{eff}}(R)}$. To this end, we use the same strategies as those described in Sec.~\ref{sec:Sampling methods}, but employ
$R$ as the only reaction coordinate in Eqs.~\eqref{eqn:Zc2}-\eqref{eqn:confentropy}.
Afterwards we proceed to the angle-resolved case.

\subsection{Angle-averaged effective potentials}\label{sec:Angle-averaged effective potentials}
\begin{figure}
\includegraphics[width=0.9\columnwidth]{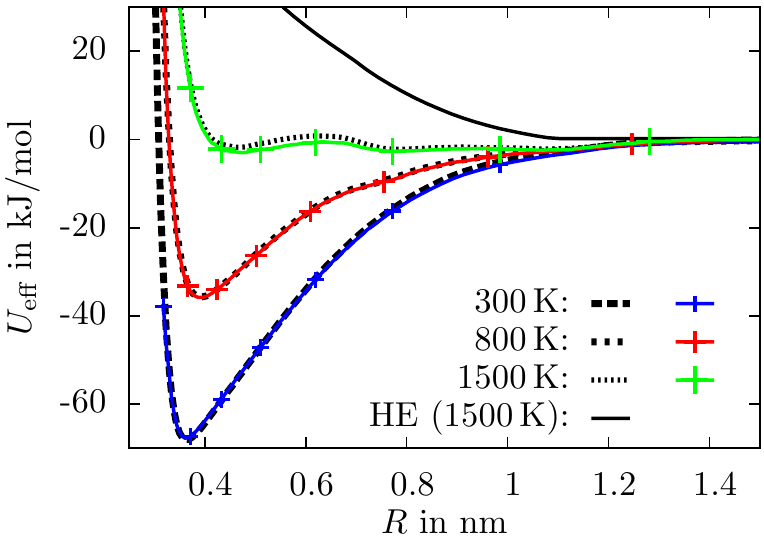}
\caption{\label{fig:angavgpots}The angle-averaged effective potential at different temperatures using steered dynamics (SD, broken lines) and umbrella sampling (US, solid colored lines).
For comparison the angle-averaged hard ellipsoidal (HE) potential for ${1500\,\mathrm{K}}$ is added.
}
\end{figure}

We calculated the effective potential  ${U_{\text{eff}}(R)}$ at three temperatures, namely ${300\,\mathrm{K}}$, ${800\,\mathrm{K}}$ and ${1500\,\mathrm{K}}$.
The angle-averaged effective potential ${U_{\text{eff}}(R)}$ is calculated in analogy to the angle-resolved effective potential (see Sec.~\ref{sec:Definition of the effective pair potential}) but with
$R$ appearing as the only reaction coordinate in Eqs.~\eqref{eqn:Zc2}-\eqref{eqn:confentropy}.
Numerical results obtained via the umbrella sampling and the steered dynamics method are presented in Fig.~\ref{fig:angavgpots}.
The data reveal a strong temperature dependence of ${U_{\text{eff}}(R)}$. At the lowest temperature considered (${300\,\mathrm{K}}$) we observe a pronounced attractive potential
well with large negative values, corresponding to a coupling strength of about $27 \, k_{\text{B}}T$.
Contrary to that, the potential at $1500\,\mathrm{K}$ is weakly positive nearly everywhere, reflecting a (weak) effective repulsion.
The pronounced temperature dependence of ${U_{\text{eff}}(R)}$ seems not too suprising in view of the amount of variables which have been integrated out. In particular,
averaging out the rotations at fixed $R$ implies that energetically most attractive configurations are mixed with less attractive ones; this mixing effect clearly becomes
the more important the higher the temperature is. In fact, in the limit of {\it infinite} temperature one would expect attractive atom-atom interactions 
to become entirely irrelevant, yielding a purely entropic effective interaction determined by only the steric repulsion between the particles. This "entropic limit"
of ${U_{\text{eff}}(R)}$ should be close to the angle-averaged potential of two hard ellipsoidal disks (HE). Numerical results for the latter are included in Fig.~\ref{fig:angavgpots}
(the data have been obtained in analogy to that between coronene molecules).
Inspecting then the temperature dependence of ${U_{\text{eff}}(R)}$ we see that, at  ${1500\,\mathrm{K}}$, we are not yet in the entropic limit but are clearly approaching it.

\par
A further interesting feature revealed by Fig.~\ref{fig:angavgpots} is that, quite independent of the temperature, the range of ${U_{\text{eff}}(R)}$ is always about ${1.2\,\mathrm{nm}}$.
Finally, we see that the two sampling methods yield numerically consistent results except for minor differences in the range ${R\approx 0.4 \text{--} 0.8 \,\mathrm{nm}}$ at ${T=1500\,\mathrm{K}}$.

\subsection{Angle-resolved effective potentials}\label{sec:Angle-resolved effective potentials}
\begin{figure*}
\begin{center}
\includegraphics[width=0.9\textwidth]{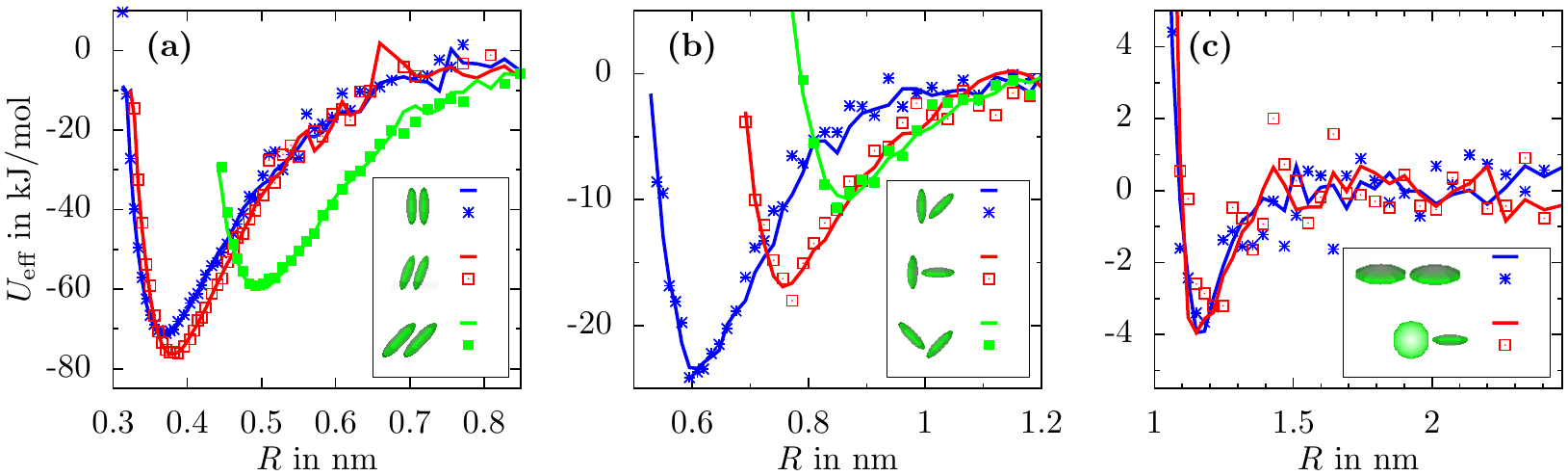}
\end{center}
\caption{\label{fig:potcurves800kmixed}
Angle-resolved effective potentials at ${T=800\,\mathrm{K}}$ obtained from umbrella sampling (solid lines) and steered dynamics (points).
(a) face-face, parallel weakly displaced and parallel displaced configuration; (b) T-, V-, and herringbone configuration; (c) edge-edge and cross configuration
}
\end{figure*}

We now turn to the central issue of this article, that is, the angle dependence of the effective potentials.
We first focus on the case ${T=800\,\mathrm{K}}$. Corresponding results for the angle-resolved potentials are shown in Fig.~\ref{fig:potcurves800kmixed}, where we concentrate on the configurations
introduced in Sec.~\ref{sec:Numerical details}.
It is seen that all configurations are characterized by an attractive well at short distances. However, the position of the potential minimum and its depth strongly
depend on the specific orientation.
The most attractive configurations are those with a large contact area of the particles, that is, the face-face and parallel-displaced configurations [see Fig.~\ref{fig:potcurves800kmixed}(a)].
Among these, the most attractive one is not the perfect face-face (as one might have expected), but the weakly parallel displaced configuration.
The corresponding potential depth is larger by about a factor of twenty as compared to that of the edge-edge and cross configuration.
Comparing now the two sampling methods, we find the US and SD method to be generally consistent, with the accuracy of each method depending somewhat on the intermolecular distance
considered. At short distances, the potential curves are better described by the SD technique, while the US method is superior (in terms of roughness) at larger distances.
Moreover, the US technique provides a particularly efficient sampling of the orientation configuration at fixed distance $R$.
To improve the performance of this latter technique at small distances one could use larger spring constants or adaptive US techniques~\cite{Mezei1987}. We also note the two methods,
US and SD, have already been compared in Ref.~\onlinecite{Bastug2008} for the case of one reaction coordinate. In that study, it was found that SD requires an order of magnitude longer simulation
runs to obtain the same accuracy as the US method.
This is consistent with our observations and can be seen for configurations with a large contact distance [see Fig.~\ref{fig:potcurves800kmixed}(c)] .
\par
\begin{figure}
\begin{center}
\includegraphics[width=0.9\columnwidth]{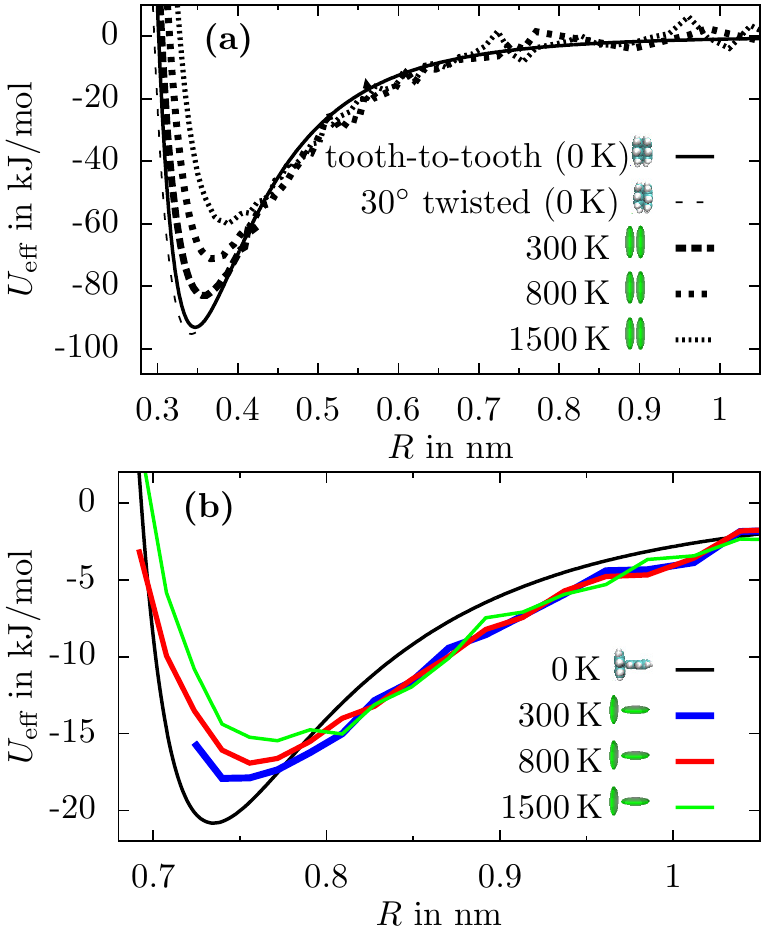}
\end{center}
\caption{\label{potcurvestemperature}
(a) Steered dynamics (SD) results for the face-face configuration at different temperatures. Included are two curves corresponding to face-face results for atomistically-resolved coronene molecules in its ground state (${T=0}$).
(b) Umbrella sampling (US) results for the T configuration at different temperatures. For comparison, a T configuration for atomistic coronene molecules in their ground state (${T=0}$) is included as well.}
\end{figure}

We next consider the influence of temperature on the angle dependent potentials. Figure~\ref{potcurvestemperature}(a)
plots as an example SD results for the face-face configuration at three temperatures. 
It is seen that there is, indeed, a temperature dependence, but this dependence is less
pronounced than in the case of the angle-averaged potential ${U_{\text{eff}}(R)}$ (see Fig.~\ref{fig:angavgpots}).
This is plausible in view of the
less severe coarse-graining: instead of integrating out {\it all} "dimer" configurations (including rotations) at fixed $R$, 
the configurational average yielding the potential in Fig.~\ref{potcurvestemperature}(a) only involves
face-face configurations with different degrees of "bending", that is, thermal fluctuations of the atoms at fixed mean orientation.
In other words, the temperature dependence is strongly correlated with the fact that, on the atomistic level, the
molecules are not rigid. 
This is also demonstrated by the difference between our finite-temperature results and those
related to {\it groundstate} configurations [included in Fig.~\ref{potcurvestemperature}(a)], where the atomistically resolved molecules are rigid.
Less temperature dependence is found for configurations with larger contact distance, such as the T-configuration [see Fig.~\ref{potcurvestemperature}(b)]. 
We understand this as a consequence of the fact that
the energy related to T-configurations is less affected by bending fluctuations. Still one clearly observes, similar to the face-face case, the tendency that a decrease of the temperature
yields a decrease of the well-depth, that is, an increase of attraction.
 \section{Parametrization in terms of a Gay-Berne potential}\label{sec:Parametrization in terms of a Gay-Berne potential}
So far we have determined the effective pair potential numerically from the atomic trajectories.
In the following, we aim at parametrizing ${U_{\text{eff}}}$ in terms of an established interacting potential, specifically a Gay-Berne (GB)-like model~\cite{Gay1981}.
The latter involves, in principle, all aspects observed in our numerical data, that is, anisotropy, softness, and attraction at short length scales. The advantage of such a parametrization
is that all potential values can be accessed without any smoothing or extrapolation. Moreover, the calculation of forces and torques is strongly simplified.
Specifically, we consider a modification of the original GB model which was introduced by Kabadi~\cite{Kabadi1986}. The modification involves a coefficient $d_{\text{w}}$ which acts
as a factor on the well width of the original GB potential.
The modified GB potential reads
\begin{align}
U(R,a,b,c)=4 \, \epsilon(a,b,c) \cdot \left[ \left(\frac{1}{R^*}\right)^{12}- \left(\frac{1}{R^*}\right)^{6} \right]\text{,}
\label{eqn:U(R,a,b,c)}
\end{align}
where ${R^*=(R-\sigma(a,b,c)+d_{\text{w}} \sigma_0)/(d_{\text{w}} \, \sigma_0})$ represents the reduced distance,  ${\sigma(a,b,c)}$ the contact function (i.e., ${U(\sigma)=0}$)  and ${\epsilon(a,b,c)}$ the well depth.
A drawback of the Gay-Berne contact distance is that for orthogonal facing particles, i.e. $c=0$, the potential becomes independent of $a$ and $b$.
We therefore employ the following definition of the contact distance~\cite{Guevara-Rodriguez2011},
\begin{align}
 \sigma(a,b,c)= \sigma_0 \left[1- \frac{\chi}{2}  \left( A^+ + A^- \right) + (1-\chi)\,\chi_{\text{t}} \left( A^+ \,A^- \right)^{\gamma} \right]^{-\frac{1}{2}}\text{.}
 \label{eqn:contact distance}
\end{align}
For the coefficients in Eq.~\eqref{eqn:contact distance}, we have $A{^{\pm}=(a\pm b)^2/(1 \pm \chi c)}$, and the anisotropy parameters
${\chi=(\kappa^2-1)/(\kappa^2+1)}$ and ${\chi_{\text{t}}=\left[(\kappa-1)/(\kappa+1)\right]^2}$, where  ${\kappa=\sigma_{\text{FF}}/\sigma_0}$ is the quotient of the face-face and edge-edge contact distance.
Regarding the well depths $\epsilon$, we use the well-known GB formula~\cite{Gay1981}
\begin{align}
 \epsilon(a,b,c)&=\epsilon_0 \cdot \left[1-\chi^2 c^2    \right]^{-\frac{1}{2}\,\nu} \cdot \left[\epsilon_{\text{M}}(a,b,c)\right]^{\mu}\text{,}
 \label{eqn:epsilon}
\end{align}
where the overlap factor $\epsilon_{\text{M}}$ is modified (as compared to the original definition~\cite{Gay1981}) according to
\begin{align}
\label{eqn:epsilonM}
\epsilon_{\text{M}}(a,b,c)&=1- \frac{\chi'}{2}  \left( A'^+ + A'^- \right) + \theta \cdot \left( A'^+ \,A'^- \right)^{\gamma'} \\
&+ \xi\cdot \left[ 1-5a-5b-15 a^2 b^2 +2\left(c-5ab\right)^2 \right]\text{.}\nonumber
\end{align}

The coefficients $A'^{\pm}=(a\pm b)^2/(1 \pm \chi' c)$ in Eq.~\eqref{eqn:epsilonM} resemble the quantities $A^{\pm}$, but incorporate the anisotropy parameter $\chi'$. 
In Eq.~\eqref{eqn:epsilonM} the term ${\theta \cdot \left( A'^+ \,A'^- \right)^{\gamma'}}$ is introduced to modify the strength of T-like configurations, while the last term
is introduced to increase the attraction strength for the parallel displaced configurations (inspired by linear quadrupole-quadrupole interactions~\cite{Stone1978,Boublik1990}).
However, it should be noted that the strength multiplicator $\epsilon_0$ does not correspond any more to the potential depth of the cross configuration.

We are now in the position to parametrize our coarse-grained potentials. In the subsequent paragraph Sec.~\ref{sec:Proposed interaction model  (M)}
we first introduce a parameter choice, 
which we call model M and which later turned out to be superior in representing various dimer configurations as compared to other parameter choices described in Sec.~\ref{sec:modelK1}. 
We also discuss the relation between the quality of the model (using the numerically coarse-grained data as a reference)
and the amount of orientational configurations used for the fitting.

\subsection{Proposed interaction model  (M)}\label{sec:Proposed interaction model  (M)}

Within model M we fix the following parameters for all temperatures: ${\mu=1}$, ${\nu=1}$, ${\gamma=4}$ and ${\gamma'=4}$.
The anisotropy parameter $\chi$ is calculated by measuring the face-face contact distance $\sigma_{\text{FF}}$, which is always sampled in our setup of steered dynamics simulations, and the edge-edge contact distance $\sigma_0$.
The well width is calculated using $\sigma_{\text{FF}}$ and the distance corresponding to the minimum of the face-face potential, $R_{\text{FF}}^{\text{min}}$, yielding
\begin{align}
 d_{\text{w}}=\frac{R_{\text{FF}}^{\text{min}}-\sigma_{\text{FF}} }{\sigma_0 \, \left( 2^{1/6} -1 \right)}\text{.}
\end{align}
To summarize, the two parameters which determine the shape, are extracted from the face-face and edge-edge configuration.
Further configurations come into play when we determine the remaining parameters $\epsilon_0$, $\chi'$, $\theta$ and $\xi$ 
by fitting the simulation results for $U_{\text{eff}}(R,a,b,c)$ according to Eq.~\eqref{eqn:epsilon}. Specifically,
our parameter fit builds on the four attractive wells stemming from the
parallel weakly displaced, parallel displaced, T and edge-edge configuration.
The main reason to pick those configurations is that, according to our many-particle simulations
presented in Sec.~\ref{sec:Many particle simulations}, these are the four most frequent configurations at high densities.
The resulting fits are presented in Figs.~\ref{fig:potfits}(a)-(j) in Appendix~\ref{sec:Effective potentials}, 
where we consider various temperatures and orientational configurations. The corresponding parameters 
at different temperatures are contained in Table~\ref{tab:M} in Appendix~\ref{sec:Parametrizations}. 
The fit curves in Figs.~\ref{fig:potfits}(a)-(j) illustrate two important features of model M: First,
although the construction of the parameter set involves only four configurations, the model gives good
results (as compared to the original coarse-grained potential) also for other configurations, such as V configuration.
Only the face-face configuration is slightly
overestimated. Second, model M is {\it intrinsically consistent} in the sense that, when performing an angle average 
over the fit results, one arrives at a potential which is very close to the angle-averaged numerical coarse-grained potential 
discussed in Sec.~\ref{sec:Angle-averaged effective potentials} (see also Sec.~\ref{sec:modelS}).

\subsection{Kabadi models (K1 and K2)}
\label{sec:modelK1}

For comparison we now introduce two further models, K1 and K2.
These are parametrizations of the Kabadi potential~\cite{Kabadi1986}, which only differs from the Gay-Berne potential through the distance-dependent part in Eq.~\eqref{eqn:U(R,a,b,c)}.
In contrast to model M, models K1 and K2 
involve only two parameters to characterize the interaction strength, namely $\epsilon_0$ and $\kappa'$. These are used as fit parameters.
The well width factor $d_{\text{w}}$ and the edge-edge contact distance $\sigma_0$ are determined like in model M, that is, through the face-face and edge-edge
configuration. Likewise, the Gay-Berne parameters $\mu$ and $\nu$ are set to $1$. In summary, models K1 and K2 are constructed
by considering two orientational configurations (rather than four as in model M) to fit the potential strength.

The remaining fitting parameters are adjusted in two different ways:
Model K1 aims at a correct reproduction of the V-configuration which is important for collisions in the isotropic phase.
Model K2 aims at a correct edge-edge well depth, which is a crucial configuration in the crystalline regime.
Both, K1 and K2, yield an accurate representation of the most attractive configuration, namely the weakly parallel displaced configuration.
However, the models are inconsistent in that they do not reproduce the coarse-grained angle-resolved potential.

\subsection{Spherical model (S)}
\label{sec:modelS}
For completeness, we also introduce a model (S) involving a pure distance-dependent potential.
To this end we use the angle-averaged potential taken from the SD simulations presented in Fig.~\ref{fig:angavgpots}.

\section{Many-particle simulations}\label{sec:Many particle simulations}

So far, we have focused on the effective potential between two coronene molecules.
In the following we aim at testing the developed coarse-grained models in the context of many-particle simulations.
To this end we compare various equilibrium properties obtained from the coarse-grained simulations with corresponding ones from atomic simulations
(see Sec.~\ref{sec:Atomic system}).

The coarse-grained (or mesoscopic) simulations are performed on the basis of Molecular Dynamics in the NVT or NPT ensemble, with P being the pressure.
The translational and rotational equations of motion are solved using the leapfrog algorithm. Temperature and pressure  control (if present)
are realized via a Berendsen thermostat (barostat)~\cite{Berendsen1984}.
In that framework, translational and rotational temperature are controlled separately.
The simulations are performed using a cutoff of ${2.4\,\mathrm{nm}}$ for the coarse-grained interactions. The time constant involved in T- or P-control is set to ${2\,\mathrm{ps}}$.
In case of pressure control we use a compressibility of ${2.25\cdot 10^{-4} \,\mathrm{bar}^{-1}}$. Finally, the moment of inertia entering the rotational equations of motion is set to
${I=14.76\,\mathrm{u\,nm}^2}$ corresponding to a coronene molecule in the ground state.
\par
We have performed mesoscopic simulations for a range of temperatures ${300\,\mathrm{K}\le T \le 1500\,\mathrm{K}}$ and different densities. For all temperatures considered, we have used the corresponding 
parametrized potentials introduced in Sec.~\ref{sec:Parametrization in terms of a Gay-Berne potential} (the temperature-dependent parameters are presented in Tables~\ref{tab:M}, ~\ref{tab:K1} and~\ref{tab:K2} in the appendix).
In the following we discuss separately representative thermodynamic states pertaining to the isotropic and orientationally ordered regime.

\subsection{Isotropic regime}
The isotropic phase was explored by using a NVT ensemble characterized by ${N=576}$, ${T=1500\,\mathrm{K}}$, and ${V_1=(15\,\mathrm{nm})^3}$ or ${V_2=(8.3\,\mathrm{nm})^3}$.
The corresponding packing fractions were ${\eta_{(1)}\approx0.04}$ and ${\eta_{(2)}\approx0.21}$, respectively. Therefor the molecules are regarded as ellipsoids of revolution, whose diameters
are taken from contact distances of $U_{\text{eff}}$ for ${T=1500\,\mathrm{K}}$. 

\begin{figure}
\includegraphics[width=0.9\columnwidth]{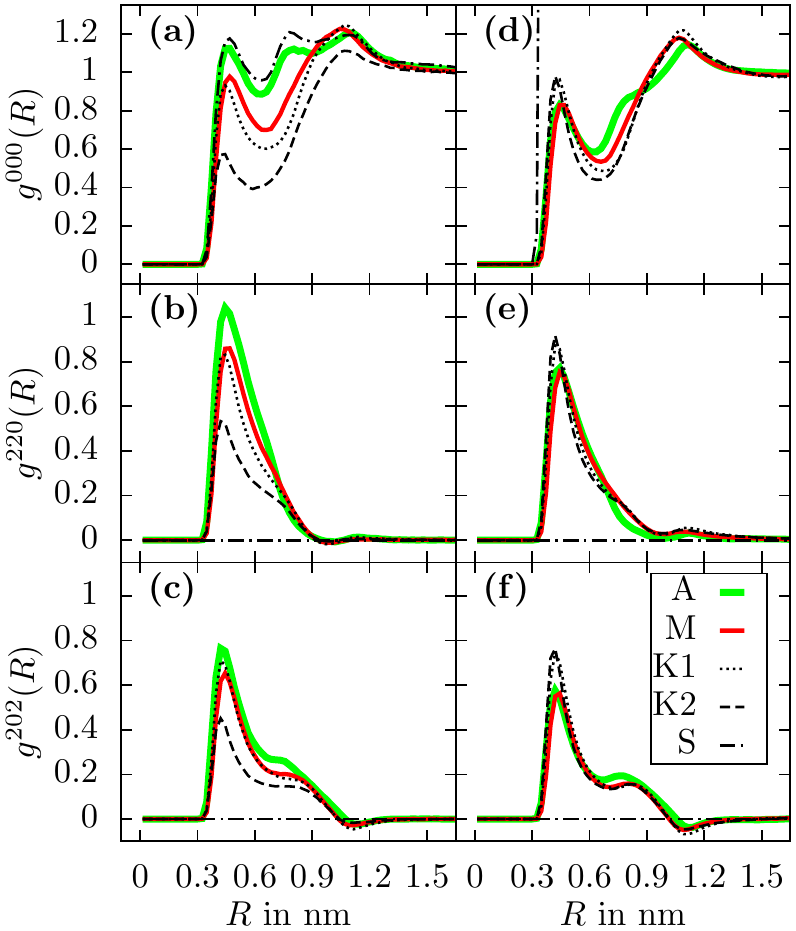}
 \caption{\label{fig:gcoefficients}
 Coefficients of the pair correlation function at a low density [parts (a)-(c)] and a higher density [parts (d)-(f)] at ${T=1500\,\mathrm{K}}$. Results are obtained using the models M, K1, K2, and the spherical model S.
 Included are data from atomic simulations (A).
 }
\end{figure}

Structural properties were extracted after an equilibration time of $2\,\mathrm{ns}$. Specifically, we considered various coefficients of the space- and orientation dependent
pair correlation function ${g(\mathbf{R},\mathbf{\hat{u}}_1,\mathbf{\hat{u}}_2)}$ in an expansion in terms of spherical invariants~\cite{Gray1984}.
The simplest one is the coefficient $g^{000}(R)$, which corresponds to the usual, angle-averaged correlation function. Further, we calculate the coefficient $g^{220}(R)$ which
involves the average of ${P_2(\mathbf{\hat{u}}_1\cdot\mathbf{\hat{u}}_2)=P_2(c)}$ with ${P_2(x)=3/2\,x^2-1/2}$. Thus, ${g^{220}(R)}$ is a measure of the mutual alignment of two molecules at a distance $R$.
Finally, we consider the function  ${g^{202}(R)}$ which involves the average of ${P_2(\mathbf{\hat{u}}_1\cdot \mathbf{R})=P_2(a)}$ and thus measures the alignment relative to the connecting vector.
Explicit statistical expressions for these functions are given in Ref.~\onlinecite{Gray1984}.
Numerical results are shown in Fig.~\ref{fig:gcoefficients}.
Considering first the lower density, we see that the angle-averaged correlation of the atomic system is best reproduced by the data from the spherical model (S).
Models M, K1, K2 underestimate the first peak in ${g^{000}(R)}$, with the largest error appearing from model K2. Regarding the non-spherical coefficients 
(for which the spherical model obviously cannot make predictions), we find that model M works best, while the largest deviation occurs again from model K2. This reflects
the fact that also the angle dependence of the K2 potential is less pronounced than in M and K1 (see Fig.~\ref{fig:potfits}).
\par
\begin{figure}[h]
\includegraphics[width=0.9\columnwidth]{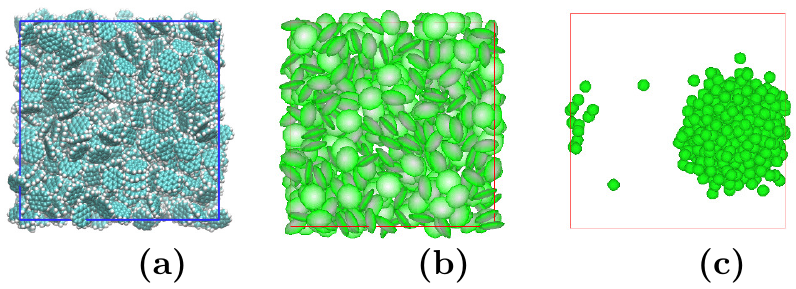}
 \caption{\label{fig:isotropicsystems}The atomistic system (a), the angle-resolved coarse-grained system (b) and the angle-averaged coarse-grained system (model M) (c) are shown at a packing fraction of ${\eta_{(2)}\approx0.21}$.}
\end{figure}

At the higher density, all three models M, K1, K2 yield good results (as compared to the atomic data) for the correlation functions considered [see Figs.~\ref{fig:gcoefficients}(d)-~\ref{fig:gcoefficients}(f)].
The best accuracy is again provided by model M. A further very interesting observation is that the model S predicts a totally unphysical result for ${g^{000}(R)}$; the latter
does not approach unity at large distances.
This aready indicates that systems characterized by this spherical potential are not in a stable equilibrium state any more.
Rather the particles condense into one big cluster, indicating a phase separation (see Fig.~\ref{fig:isotropicsystems}). 
This phenomenon, which is absent in the atomic system, clearly indicates that modeling the system with an angle-averaged
potential is not appropriate, at least not at intermediate and high densities. We take this failure as an a-posteriori justification for our effort to obtain angle-dependent potentials.
\par
As a further test of our potentials, we have calculated the virial pressure~\cite{Allen1989} and the second virial coefficient, $B_2$.
The results are summarized in Table~\ref{tab:pressure}.
At the lower density $\eta_{(1)}$, the pressure values predicted by the various models are fairly similar, and the second virial coefficient is rather small.
This indicates that the pressure is dominated by its ideal-gas value. We also see that (at $\eta_{(1)}$) model S is closest to the atomic value, consistent with the corresponding
observation for ${g^{000}(R)}$ (see Fig.~\ref{fig:gcoefficients}). At the larger density $\eta_{(2)}$ the differences in the pressure data are larger, as expected.
The closest match of the atomic value is given by model M [again consistent with our previous discussion of ${g^{000}(R)}$]. We also see that the pressure predicted by the spherical model is too small
by two orders of magnitude. This is just another manifestation of the above-mentioned failure of this model to predict a stable liquid phase. We note, however, that the corresponding value of $B_2$
matches per definition that of the atomic system, due to the fact that $B_2$ is solely a function of the two-particle configuration integral, $Z_2^{\text{c}}$ and the volume $V$ according to
\begin{align}
 \label{eqn:B2}
 B_2=\frac{V}{2}-\frac{Z_2^{\text{c}}\,\alpha^6}{V}\text{,}
\end{align}
which are both fixed during coarse-graining (see Sec.~\ref{sec:Definition of the effective pair potential}).
\begingroup
\squeezetable
\begin{table}[h]
\begin{ruledtabular}
\caption{\label{tab:pressure}
Atomic and coarse-grained simulation results for the pressure $P$ and the pressure correction due to the second virial coefficient $B_2$
are shown for the two different system densities $\rho_{(1)}=576/(15\, \mathrm{nm})^3$ and $\rho_{(2)}=576/(8.3\, \mathrm{nm})^3$, respectively, at ${T=1500\,\mathrm{K}}$.
The second virial coefficient of the atomic system matches per definition that of model S (see main text).
}
\begin{tabular}{ccccc}
Model&$P_{(1)}$ $[\mathrm{bar}]$& $B_2\, \rho_{(1)}^2/\beta$ $[\mathrm{bar}]$& $P_{(2)}$ $[\mathrm{bar}]$ & $B_2\, \rho_{(2)}^2/\beta$ $[\mathrm{bar}]$\\
\hline
&&&&\\
atomic (A)&31.17& see S& 200.21 & see S \\ 
 M&33.36&-0.74& 246.29 & -25.91\\ 
 K1&34.01&0.07& 260.03 & 2.40\\ 
 K2&38.75&4.99& 426.72 & 173.76\\ 
 S&30.70&-0.84& 1.71 & -29.38\\ 
ideal gas&35.35&0& 208.63 & 0\\ 
\end{tabular}
\end{ruledtabular}
\end{table}
\endgroup
\subsection{Columnar hexagonal nematic regime}

\begin{figure}[h]
\includegraphics[width=0.9\columnwidth]{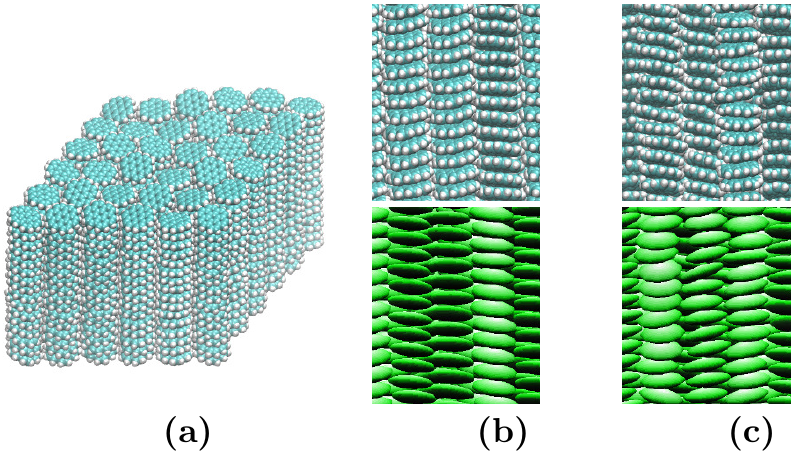}
 \caption{\label{fig:crystal}
(a) Initial hexagonal columnar nematic state of the present system.
(b) and (c): Snapshots of the hexagonal nematic regime after ${3\,\mathrm{ns}}$ at b) ${300\,\mathrm{K}}$ and (c) ${700\,\mathrm{K}}$ within the atomic model and model M.}
\end{figure}

At lower temperatures and sufficiently high densities the present coronene system displays orientationally ordered phases on both, the atomic and the coarse-grained level. One of these phases is 
a columnar phase characterized by a nematic ordering of the molecular symmetry axes and a hexagonal arrangement of the columns in the plane perpendicular to the column axes.
The same type of phase also occurs in conventional Gay-Berne systems consisting of diskotic particles~\cite{Emerson1994}.
Moreover, columnar nematic phases have also been observed in systems of hexabenzocoronene derivatives~\cite{Andrienko2006}.
In the following we investigate the stability of this high-density phase in both, the atomic and the coarse-grained simulations. 
To this end we performed constant-pressure simulations.
In order to initialize the simulations, we first set up a {\em perfect} hexagonal columnar configuration,
involving only face-face and edge-edge configurations with nearest-neighbor distances of $0.37$ and $1.13\,\mathrm{nm}$, and then 
applied a Gromacs energy minimization routine (steepest descent method) [see Fig.~\ref{fig:crystal}(a)] yielding the starting configuration for our simulation.
Simulations have then been performed for temperatures ranging from ${300\,\mathrm{K}}$ to ${1500\,\mathrm{K}}$ in steps of ${100\,\mathrm{K}}$, leading from the orientationally ordered into an isotropic regime.
In all stages of these ``melting'' simulations, the pressure was fixed at $1\,\mathrm{bar}$ (with a compressibility five times larger than that of water), and the box shape (parallelepiped) was allowed
to change its geometry (see Ref.~\onlinecite{Berendsen1984}). The equilibration time varied between ${3\,\mathrm{ns}}$ (ordered regime) and ${95\,\mathrm{ns}}$ (isotropic regime).

\begin{figure*}
\begin{center}
\includegraphics[width=0.9\textwidth]{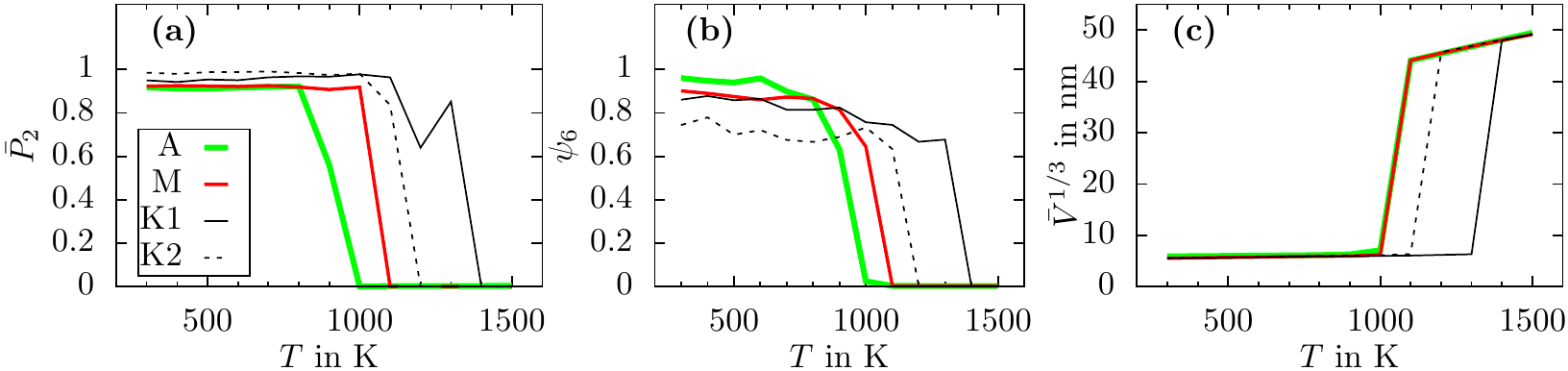}
\end{center}
 \caption{\label{fig:ops}Order parameters [see Eqs.~\eqref{eqn:P2} and ~\eqref{eqn:Psi6}] and characteristic length obtained from the average volume as functions of temperature and different models.}
\end{figure*}

To evaluate the overall degree of ordering we calculated the nematic order parameter $\bar{P}_2$ defined by
\begin{align}
 \bar{P}_2=\frac{1}{N}\left< \sum_i P_2(\mathbf{\hat{u}}_i\cdot \mathbf{\hat{n}}) \right>\text{,}
 \label{eqn:P2}
\end{align}
where $\mathbf{\hat{n}}$ is a unit vector indicating the direction of the nematic director.
The latter was taken to be the surface normal along which the columns are set up.
Further, we consider a hexagonal bond order parameter ($\Psi_6$) suitable for columnar configurations. The latter is defined as

\begin{align}
\psi_6=\frac{1}{N}\left<\sum_j\sum_k^{Nb_j} e^{\mathrm{i}\, 6\,\theta_{jk}}\right>\text{,}
\label{eqn:Psi6}
\end{align}

where $Nb_j$ is the number of neighbors of molecule $j$.
Here, particles are considered neighbors if the projection of the connecting vector ${\mathbf{R}_k-\mathbf{R}_j}$ onto $\mathbf{\hat{n}}$ is smaller than $0.25\,\mathrm{nm}$, while the projection perpendicular
to $\mathbf{\hat{n}}$ is between ${0.9\,\mathrm{nm}}$ and ${1.3\,\mathrm{nm}}$.
The behavior of these two order parameters, as well as that of the calculated volume as functions of temperature is plotted in Fig.~\ref{fig:ops}, where we have included results
from the atomic system (A) and from the coarse-grained models M, K1, K2.
At ${300\,\mathrm{K}}$, the atomic system displays nearly perfect nematic and columnar ordering, that is both order parameters are close to unity.
Upon increasing $T$, $\bar{P}_2$ (of the atomic system) exhibits a sudden decrease at ${T\approx 900\,\mathrm{K}}$, indicating the disappearance of nematic ordering.
The parameter $\psi_6$ also decreases, but in a smoother way. This reflects the observation that the columns first somewhat rearrange before the columnar structure finally melts.
As a consequence of melting, the volume of the system strongly increases, as indicated by the plot of the third root of the average volume in Fig.~\ref{fig:ops}(c). 
All of the coarse-grained models M, K1, K2 reproduce qualitatively the phase transition of the atomic system, although the predicted transition temperatures are clearly model-dependent.
Taken altogether, model M provides the best representation of the atomic data. Although the drop of $\bar{P}_2$ upon heating occurs at a somewhat too high temperature, the characteristic length provided by the third
root of the volume is reproduced very accurately. Compared to M [and the atomic system (A)], we find that the first Kabadi model (K1) predicts the columnar melting at significantly larger
temperatures (${T=1300\,\mathrm{K}-1400\,\mathrm{K}}$). This can be explained by the fact that model K1 overestimates the attraction associated with the edge-edge configuration. Finally, the results of model K2
(which gives a correct edge-edge configuration), are in between those of models M and K1.
\par
\begin{figure}[h]
\includegraphics[width=0.9\columnwidth]{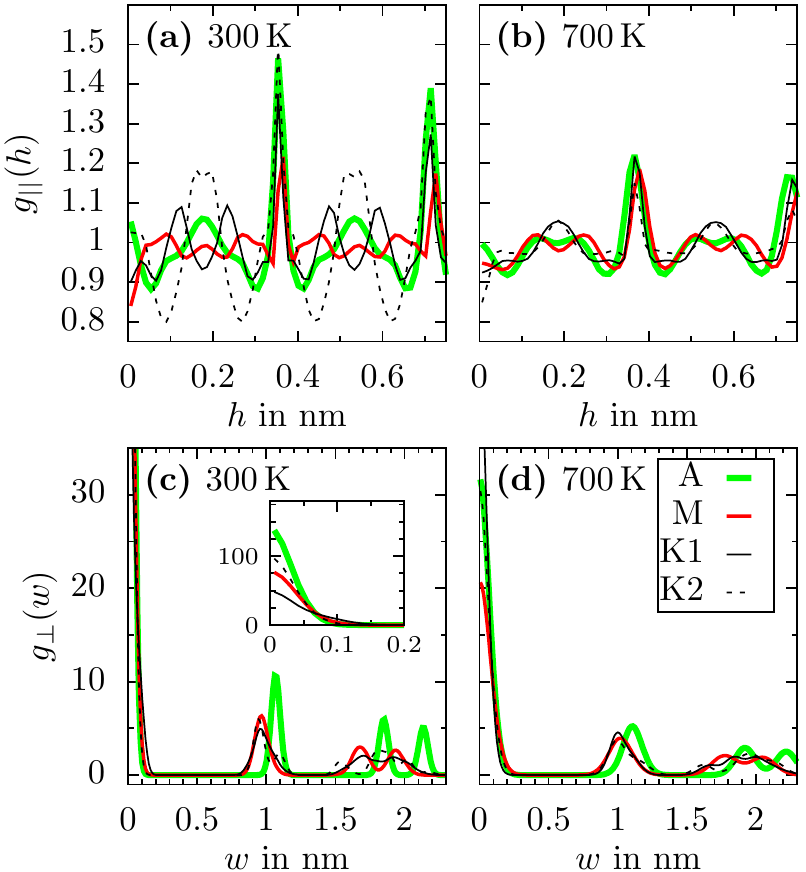}
\caption{\label{fig:distrfcts}
Pair distribution functions along the director ($g_{||}$) and perpendicular to it ($g_{\bot}$) at ${T=300\,\mathrm{K}}$ and ${T=700\,\mathrm{K}}$ for the atomic (A) and the coarse-grained model systems (M, K1, K2).
}
\end{figure}

In addition to the system-averaged order parameters discussed so far, we have also investigated the local structure in the columnar nematic phase. To this end we consider the correlation
functions parallel and perpendicular to the nematic director $\mathbf{\hat{n}}$, $g_{\bot}$ and $g_{||}$.
These functions are calculated based on expressions suggested in Ref.~\onlinecite{Andrienko2006}. However, here we consider normalized versions (where the correlation functions yield unity
if no correlation is present). Specifically, 

\begin{subequations}
\begin{align}
 g_{||}(h)&=\left<\frac{2}{V_{||}\,N\, \rho} \sum_{j}\sum_{k>j} f(h,\mathbf{\hat{n}}\cdot \mathbf{R}_{jk})\right>\text{,}\label{eqn:gpoa}\\
 g_{\bot}(w)&=\left<\frac{2}{V_{\bot}(w)\,N\, \rho} \sum_{j}\sum_{k>j} f(w,\left|\mathbf{\hat{n}}\times \mathbf{R}_{jk}\right|)\right>\label{eqn:gpob}\text{,}
\end{align}
\end{subequations}

where ${f(x,y)}$ equals unity for ${y \in \left( x -\frac{\Delta}{2}, x +\frac{\Delta}{2} \right]}$, otherwise ${f=0}$ (with $\Delta$ being the bin size).
Further, the volumes appearing in Eqs.~\eqref{eqn:gpoa} and ~\eqref{eqn:gpob} are defined as ${V_{||}=\Delta\, \left| \mathbf{L}_1 \times \mathbf{L}_2 \right|}$ and ${V_{\bot}(w)=\pi\Delta\,(2w+\Delta)\,V \left| \mathbf{L}_1 \times \mathbf{L}_2 \right|^{-1}}$,
respectively, where $\mathbf{L}_1$ and $\mathbf{L}_2$ are vectors along the sides of the simulation cell perpendicular to the director.
Numerical results for the correlation functions are plotted in Fig.~\ref{fig:distrfcts}, where we consider two temperatures within the nematic columnar regime.
The atomic results for the function $g_{||}$ at ${300\,\mathrm{K}}$ [see Fig.~\ref{fig:distrfcts}(a)] clearly signal the preferred layer separation of ${0.33\,\mathrm{nm}}$ by sharp peaks.
However, one also notices a secondary, weaker maximum at intermediate distances. These latter maxima indicate that a few columns are shifted along one another by half the thickness of
a molecule. Considering the corresponding coarse-grained results we see that model K2 reproduces not only the main peaks of $g_{||}$ at ${300\,\mathrm{K}}$, but even overestimates the intermediate ones.
This results from the fact that K2 strongly favors face-face configurations relative to T- and V-like configurations.
The other models (M, K1) generate a somewhat less rigid structure. At ${700\,\mathrm{K}}$ all coarse-grained models reproduce the features seen in the atomic data for $g_{||}$ [see Fig.~\ref{fig:distrfcts}(b)].
In particular, compared to ${300\,\mathrm{K}}$, the data at ${700\,\mathrm{K}}$ reveal that the layer-to-layer distance has slightly increased.
\par
Regarding the perpendicular correlations within the columnar phase, we find that model M yields the best reproduction of the hexagonal column arrangement in the atomic system;
however, the lateral column separation is somewhat too small (as it is for models K1 and K2). This holds for both temperatures considered. The main temperature effect in both, $g_{||}$ and 
$g_{\bot}$, consists of a widening of peaks.

This is also indicated by direct inspection
of the simulation snapshots presented in Figs. ~\ref{fig:crystal}(b) and ~\ref{fig:crystal}(c).
At ${300\,\mathrm{K}}$, the dominating structure is a face-face configuration in a tooth-to-tooth setup. On the contrary, at ${700\,\mathrm{K}}$ tooth-to-tooth-like configurations have essentially disappeared.
The molecules rather seem to rotate freely around their symmetry axis.
\section{Conclusion} \label{sec:Conclusion}
In this paper we proposed an approach for calculating a distance- and angle-dependent effective pair potential between uniaxial molecules. Following
Kirkwood's route, the potential is defined as the difference between free energy profiles, where the usual dependence on the
(center of mass) distance is supplemented by additional ``reaction coordinates'' describing the molecule's relative orientation.

To extract the required information from underlying all-atom simulations we used two sampling methods, namely umbrella sampling and steered dynamics.
Within the steered dynamics method (see Sec.~\ref{sec:Steered dynamics}), we introduced a factorization of the unconstrained histogram function into two terms.
Thereby the distance-dependent part is treated conventionally, while the remaining part is determined by recording histograms of the orientational reaction coordinates ($a,b,c$) 
at each distance $R$.

As a benchmark system we have considered a pair of coronene-like molecules, 
using the generalized Amber force field~\citep{Wang2004} (without electrostatic contributions) to describe the
atomistic Hamiltonian. The resulting coarse-grained potentials reveal a strong angle- and temperature dependence.
Regarding the sampling method, we find the US and SD method to be generally consistent, with the accuracy of each method depending somewhat on the intermolecular distance
considered.

In a further step we have fitted the coarse-grained potentials onto variants of Gay-Berne-models. Thus, we have provided fit parameters for
two Kabadi~\cite{Kabadi1986} models (K1, K2), an own extension of the Gay-Berne~\cite{Gay1981} model (M), and (as a critical test) an 
angle-averaged model (S). The quality of the resulting models has been evaluated
by comparing the resulting many-particle behavior at different thermodynamic state points with that of the underlying atomistic system.

Model~M was found to be superior in most aspects, including the 
description of the orientational phase transition occuring at high densities.
However, it is also the most complex model in terms of the number of parameters involved.
Indeed, the strength parametrization involves four orientational dimer configurations,
which have been chosen due to their relevance under strongly coupled conditions.
The performance of the simpler Kabadi models, K1 and K2, (whose potential strength parametrizations are based 
on only two configurations) depends on the state considered. Specifically,
K1 gives reliable results for the isotropic phase, while K2 works better with respect to 
the columnar--isotropic transition. Another important finding is that the most 
simple, spherical model (S) is useful only in the strongly diluted isotropic phase. Besides the obvious
incapability of this model to predict orientationally ordered phases it 
falsely predicts, already at intermediate densities, a condensation transition, which is absent when using the angle-resolved (and atomistic) potentials.
These findings clearly justify the enhanced effort in determining angle-resolved potentials.
Morevover, to properly describe the various phases it turned out to be crucial to take into account the
pronounced temperature dependence of the angle-dependent potential. An important example is the effective interaction
in the face-face configuration, which is strongly affected by bending fluctuations stemming from the molecule's non-rigidity on the atomistic level.

We also note that, when comparing the computational time per core of the mesoscopic simulations
(which were based on a self-written code), on the one hand, 
and the atomistic simulation time (based on the
GROMACS package, version 4.5.5), one the other hand, we reached a speedup of about a factor 3 to 8. 
Part of this speed-up is likely due to the fact that we have 
represented the coronene molecules by a particularly simple shape, namely an uniaxial disk.
The performance of other representations, e.g. a collection of fused rings \cite{Obolensky2007} remains to be explored.

For true coronene (or other conjugated molecules) one drawback of our study clearly is that
we did not take into account electrostatic effects. We note again that full treatment of this problem would involve
to take into account not only static charges (which could, e.g., been estimated by the groundstate energy values given in Ref.~\onlinecite{Zhao2008}),
but also polarizability effects. These issues were beyond the scope of the present paper. 
On a coarse-grained level, one intuitive starting point to include electrostatics, yet without polarizability, would be to include
a quadrupole moment. The latter corresponds to the lowest-order multipole moment of coronene.
There exist already some simulation studies involving diskotics with quadrupole moments
\cite{Bates1998,Orlandi2007}, including applications to benzene \cite{Golubkov2006} (represented by a Gay-Berne disk with a linear quadrupole moment) and 
micron-sized colloidal (e.g., clay) particles \cite{Dijkstra1995,Trizac2002}. Inspired by these studies we 
currently work on an extentension of the present approach, where model~M
is supplemented by the interaction of two linear quadrupoles chosen along the symmetry axes of the particles. Our results will be reported
in a forthcoming paper.\begin{acknowledgments}
This work was supported by the Deutsche Forschungsgemeinschaft within the framework of the CRC 951 (projects A7 and A1).
\end{acknowledgments}

\appendix
\section{Configurational entropy for (R,a,b,c)}\label{sec:Configurational entropy for (R,a,b,c)}
In this appendix we calculate the quantity ${{\mathcal Z}_{\text{mic}}(R,a,b,c)}$ and thereby the related configurational entropy introduced in Sec.~\ref{sec:Definition of the effective pair potential}.
To this end we make use of the mapping functions introduced in Sec.~\ref{sec:Reaction coordinates}.
First, the microscopic configuration integral is defined as
\begin{widetext}
\begin{align}\label{eqn:Zmic1}
 {\mathcal Z}_{\text{mic}}(R,a,b,c)=\frac{1}{\alpha^{6N-1}}\int_{V} \!\! d\mathbf{r}_1\dots\! \!\int_{V} \! d\mathbf{r}_{2N}  \,\delta(R\!-\!\tilde{R}(\{\mathbf{r}\}))\dots\delta\left(c\!-\!\tilde{c}(\{\mathbf{r}\})\right)\text{.}
\end{align}
In the following, the functions labeled with circles (i.e. $\mathring{\mathbf{R}}_{\text{A}}$, $\mathring{\mathbf{R}}_{\text{B}}$, $\mathring{\hat{\mathbf{u}}}_{\text{A}}$, $\mathring{\hat{\mathbf{u}}}_{\text{B}}$) map atomic coordinates on corresponding position and orientation vectors of each molecule, while functions with a bar (i.e. $\bar{R}$, $\bar{a}$ $\bar{b}$ $\bar{c}$) map those vectors onto the reaction coordinates $R$, $a$, $b$ and $c$. 
By using identities such as ${\int_V d\mathbf{R}_{\text{A}} \,\delta(\mathbf{R}_{\text{A}}-\mathring{\mathbf{R}}_{\text{A}}(\{\mathbf{r}\})) =1}$ and  expressions like $\tilde{R}(\{\mathbf{r}\})=\bar{R}(\mathring{\mathbf{R}}_{\text{A}}(\{\mathbf{r}\}),\mathring{\mathbf{R}}_{\text{B}}(\{\mathbf{r}\}))$,
Eq.~\eqref{eqn:Zmic1} reads,
\begin{multline}\label{eqn:Zmic3}
{\mathcal Z}_{\text{mic}}(R,a,b,c)=
\\
\frac{1}{\alpha^{6N-1}}\int_V d\mathbf{R}_{\text{A}} \int_V d\mathbf{R}_{\text{B}} \int_{S^2} d\mathbf{\hat{u}}_{\text{A}} \int_{S^2} d\mathbf{\hat{u}}_{\text{B}}      
  \,\delta(R\!-\!\bar{R}(\mathbf{R}_{\text{A}}, \mathbf{R}_{\text{B}}))\dots\delta\left(c\!-\!\bar{c}(\mathbf{R}_{\text{A}}, \mathbf{R}_{\text{B}}, \mathbf{\hat{u}}_{\text{A}}, \mathbf{\hat{u}}_{\text{B}})\right)   
  \\
  \int_{V} \!\! d\mathbf{r}_1\dots\! \!\int_{V} \! d\mathbf{r}_{2N}
\delta(\mathbf{R}_{\text{A}}-\mathring{\mathbf{R}}_{\text{A}}(\{\mathbf{r}\})) \, \delta(\mathbf{R}_{\text{B}}-\mathring{\mathbf{R}}_{\text{B}}(\{\mathbf{r}\}))\,
 \delta(\mathbf{\hat{u}}_{\text{A}}-\mathring{\hat{\mathbf{u}}}_{\text{A}}(\{\mathbf{r}\})) \, \delta(\mathbf{\hat{u}}_{\text{B}}-\mathring{\hat{\mathbf{u}}}_{\text{B}}(\{\mathbf{r}\}))\text{.}
\end{multline}
The expression in the last line is denoted by ${X(\mathbf{R}_{\text{A}},\mathbf{R}_{\text{B}} ,\mathbf{\hat{u}}_{\text{A}},\mathbf{\hat{u}}_{\text{B}})}$.
We split this integral with respect to the two molecules, that is
\begin{multline}\label{eqn:XRRuu}
X(\mathbf{R}_{\text{A}},\mathbf{R}_{\text{B}} ,\mathbf{\hat{u}}_{\text{A}},\mathbf{\hat{u}}_{\text{B}})=
\left(\int_{V} \!\! d\mathbf{r}_1\dots\! \!\int_{V} \! d\mathbf{r}_{N}
\delta(\mathbf{R}_{\text{A}}-\mathring{\mathbf{R}}_{\text{A}}(\{\mathbf{r^{\text{A}}}\})) \, \delta(\mathbf{\hat{u}}_{\text{A}}-\mathring{\hat{\mathbf{u}}}_{\text{A}}(\{\mathbf{r^{\text{A}}}\}))\right)
\\
\times
\left(\int_{V} \!\! d\mathbf{r}_{N+1}\dots\! \!\int_{V} \! d\mathbf{r}_{2N}
 \delta(\mathbf{R}_{\text{B}}-\mathring{\mathbf{R}}_{\text{B}}(\{\mathbf{r^{\text{B}}}\}))  \, \delta(\mathbf{\hat{u}}_{\text{B}}-\mathring{\hat{\mathbf{u}}}_{\text{B}}(\{\mathbf{r^{\text{B}}}\}))\right)\text{,}
\end{multline}
where ${\{\mathbf{r^{\text{A}}}\}=(\mathbf{r}_1,\dots,\,\mathbf{r}_N, \mathbf{0},\dots,\mathbf{0})}$ and ${\{\mathbf{r^{\text{B}}}\}=(\mathbf{0},\dots,\mathbf{0},\mathbf{r}_{N+1},\dots,\,\mathbf{r}_{2N})}$.
The first factor symbolizes a measure for the number of microscopic realizations for a molecule A at $\mathbf{R}_{\text{A}}$ with orientation $\mathbf{\hat{u}}_{\text{A}}$.
This measure is invariant concerning the position and orientation of molecule A.
The second factor in Eq.~\eqref{eqn:XRRuu} can be treated analogously for molecule B.
As a result, ${X(\mathbf{R}_{\text{A}},\mathbf{R}_{\text{B}} ,\mathbf{\hat{u}}_{\text{A}},\mathbf{\hat{u}}_{\text{B}})=:X}$ is constant. 
Therefore Eq.~\eqref{eqn:Zmic3} can be simplified to
\begin{multline}\label{eqn:Zmic4}
 {\mathcal Z}_{\text{mic}} 
 =\frac{1}{\alpha^{6N-1}}\,X \, \int_V d\mathbf{R}_{\text{A}} \int_V d\mathbf{R}_{\text{B}} \, \delta(R\!-\!\bar{R}(\mathbf{R}_{\text{A}}, \mathbf{R}_{\text{B}})) \\
 \int_{S^2} d\mathbf{\hat{u}}_{\text{A}} \int_{S^2} d\mathbf{\hat{u}}_{\text{B}}      
  \,\delta\left(a\!-\!\bar{a}(\mathbf{R}_{\text{A}}, \mathbf{R}_{\text{B}}, \mathbf{\hat{u}}_{\text{A}})\right)\dots\delta\left(c\!-\!\bar{c}(\mathbf{R}_{\text{A}}, \mathbf{R}_{\text{B}}, \mathbf{\hat{u}}_{\text{A}}, \mathbf{\hat{u}}_{\text{B}})\right)\text{.}
\end{multline}	
By using translational and rotational invariance of the molecular dimer system, then it follows
\begin{align}\label{eqn:Zmic5}
 {\mathcal Z}_{\text{mic}}(R,a,b,c)=\frac{1}{\alpha^{6N-1}}\,X \,V\,4\pi\,R^2  \int_{S^2} d\mathbf{\hat{u}}_{\text{A}} \int_{S^2} d\mathbf{\hat{u}}_{\text{B}}  \,\delta\left(a\!-\!\bar{a}(\mathbf{R}_{\text{A}}, \mathbf{R}_{\text{B}}, \mathbf{\hat{u}}_{\text{A}})\right)\dots\delta\left(c\!-\!\bar{c}(\mathbf{R}_{\text{A}}, \mathbf{R}_{\text{B}}, \mathbf{\hat{u}}_{\text{A}}, \mathbf{\hat{u}}_{\text{B}})\right)\text{.}
\end{align}
Finally, the integration over different orientations $\mathbf{\hat{u}}_{\text{A}}$, $\mathbf{\hat{u}}_{\text{B}}$ can be written in polar coordinates with polar angles $\theta_1$ and $\theta_2$ and azimuthal angles $\phi_1$ and $\phi_2$.
Furthermore by taking into account chirality invariance, i.e. ${\phi_2\rightarrow 2\phi_1-\phi_2}$, we can define ${\psi=\left|\phi_1-\phi_2\right|}$ instead of $\phi_1$ and $\phi_2$. We find
\begin{multline}\label{eqn:Zmic6}
{\mathcal Z}_{\text{mic}}(R,a,b,c)= \frac{1}{\alpha^{6N-1}}\,X\,V\, R^2\,64\pi^2\\
\int_0^{\frac{\pi}{2}}  d \theta_1  \, \sin(\theta_1)\,  \delta(a-\cos(\theta_1)) \,
\int_0^{\frac{\pi}{2}}   d \theta_2 \, \sin(\theta_2)   \, \delta(b-\cos(\theta_2))  \\
\int_0^{\pi} d\psi\, \delta(c-(\sin(\theta_2)\cos(\psi)+\cos(\theta_1)\, \cos(\theta_2)))\text{.}
\end{multline}
From Eq.~\eqref{eqn:Zmic6}, the configurational entropy follows immediately with Eq.~\eqref{eqn:confentropy}.
\end{widetext}
\section{The WHAM equations}\label{sec:The WHAM equations}
WHAM equations~\cite{Kumar1992} serve to calculate unbiased histograms $P$ from biased histograms $P_k^{\text{bias}}$ [see Sec.~\ref{sec:Umbrella sampling}].
They consist of the following self-consistent set of equations (here shown for the reaction coordinate $R$ alone)
\begin{subequations}\label{eqn:WHAMoriginal}
\begin{align}
 \label{eqn:WHAMoriginala}
P(R)&=\sum_{k=1}^{N_w} \gamma_k(R) \, P_k^{\text{bias}}(R)\\
\label{eqn:WHAMoriginalb}
\gamma_k(R)&=\frac{n_k} { \sum_{i=1}^{N_w} n_i \, e^{-\beta (V_i(R)-F_i)}  }\\
\label{eqn:WHAMoriginalc}
F_k&=-\frac{1}{\beta} \ln\left[\int_{0}^l\!\!\! dR \,e^{-\beta V_k(R)} P(R) \right]\text{.}
\end{align}
\end{subequations}
In Eq.~\eqref{eqn:WHAMoriginalb}, $n_k$ represents the number of sampling points in umbrella window k.
Furthermore, the weights ${\gamma_k(R)}$ represent the coefficients in the decomposition of the unbiased histogram into the biased ones [see Eq.~\eqref{eqn:WHAMoriginala}].
Finally, $F_k$ are the free energy constants, which are initialized by the noniterative free energy perturbation method~\cite{Haydock1990,Roux1995}.
\begin{widetext}
\section{Dimer configurations}\label{sec:Dimer configurations}

\begingroup
\squeezetable
\begin{table}[H]
\tiny
\begin{ruledtabular}
\caption{\label{tab:tolerances}
This table summarizes the tolerances used to define specific dimer configurations, which are par\-ti\-cu\-lar\-ly relevant. Note that each configuration remains
its character by swapping the particles, which can be done by swapping $a$ and $b$.
The herringbone configuration is taken from Ref.~\onlinecite{Echigo2007}.
}
\begin{tabular}{r|ccc|ccc|ccc|ccc|ccc|ccc|ccc|ccc}
&\multicolumn{3}{c|}{face-face}&\multicolumn{3}{c|}{parallel weakly}&\multicolumn{3}{c|}{parallel displaced}&\multicolumn{3}{c|}{T}&\multicolumn{3}{c|}{herringbone}
&\multicolumn{3}{c|}{V}&\multicolumn{3}{c|}{edge-edge}&\multicolumn{3}{c}{cross}\\
&\multicolumn{3}{c|}{\includegraphics[height=1.8em]{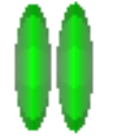}}&\multicolumn{3}{c|}{displaced \includegraphics[height=1.8em]{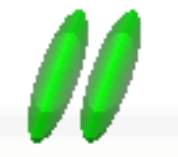}}&
\multicolumn{3}{c|}{\includegraphics[height=1.8em]{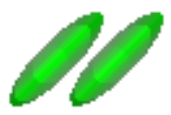}}&\multicolumn{3}{c|}{\includegraphics[height=1.8em]{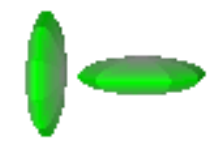}}&
\multicolumn{3}{c|}{\includegraphics[height=1.8em]{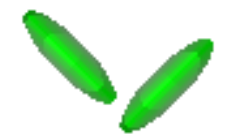}}&\multicolumn{3}{c|}{\includegraphics[height=1.8em]{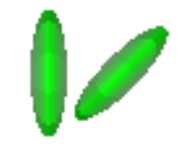}}&
\multicolumn{3}{c|}{\includegraphics[height=1.8em]{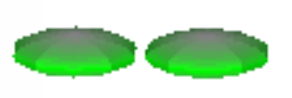}}&\multicolumn{3}{c}{\includegraphics[height=1.8em]{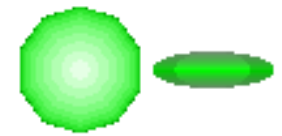}}\\
& a & b & c& a & b & c& a & b & c& a & b & c& a & b & c& a & b & c& a & b & c& a & b & c\\
\hline
&&&&&&&&&&&&&&&&&&&&&&&&\\
value & $1$ & $1$ & $1$ & $0.94$ & $0.94$ & $1$ & $1/\sqrt{2}$ & $1/\sqrt{2}$ & $1$ & $0$ & $1$ & $0$ & $0.8767$ & $0.481$& $0$ & $1$ & $1/\sqrt{2}$ & $1/\sqrt{2}$& $0$ & $0$ & $1$& $0$ & $0$ & $0$\\
min & $0.96$ & $0.96$ & $-1$ & $0.92$ & $0.92$ & $0.96$ & $0.68$ & $0.68$ & $0.96$ & $0$ & $0.88$ & $-0.12$ & $0.8$ & $0.44$ & $-0.12$ & $0.88$ & $0.64$ & $0.64$ & $0$ & $0$ & $0.8$ & $0$ & $0$ & $-0.2$\\
max & $1$ & $1$ & $1$ & $0.96$ & $0.96$ & $1$ & $0.72$ & $0.72$ & $1$ & $0.12$ & $1$ & $0.12$ & $0.92$ & $0.56$ & $0.12$ & $1$ & $0.76$ & $0.76$ & $0.2$ & $0.2$ & $1$ & $0.2$ & $0.2$ & $0.2$
\end{tabular}
\end{ruledtabular}
\end{table}
\endgroup
\vfill*

\section{Effective potentials}\label{sec:Effective potentials}

\begin{figure}[H]
\begin{center}
\includegraphics[width=0.85\columnwidth]{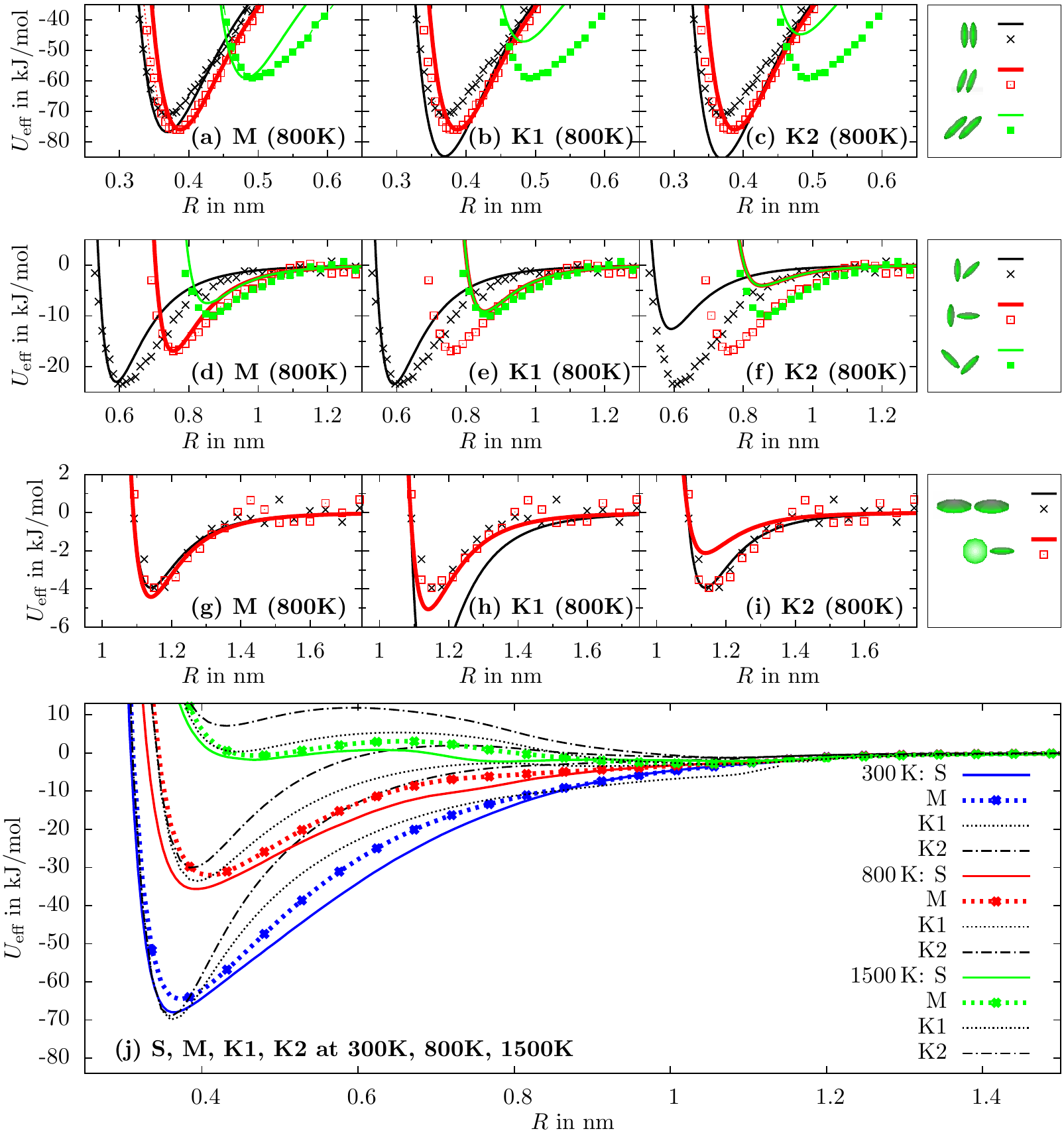}
\end{center}
 \caption{\label{fig:potfits}
(a) - (i) Effective potentials for selected configurations and corresponding fits at ${800\,\mathrm{K}}$. In each subfigure the dots represent the corresponding reference curves.
Reference curves in the top row [(a), (b), (c)] stem from steered dynamics, while the remaining reference curves stem from umbrella sampling calculations.
The left column shows the performance of our proposed model (M).
The two right columns represent the fits according to the Kabadi models K1 and K2. 
Fig.~(j) displays angle-averaged effective potentials for all considered models at ${300\,\mathrm{K}}$, ${800\,\mathrm{K}}$ and ${1500\,\mathrm{K}}$.
Model S (angle-averaged curve from steered dynamics) provides reference curves.
}
\end{figure}


\section{Parametrizations}\label{sec:Parametrizations}
The following tables summarize the parameters used to fit the effective potentials according to models M, K1, K2 for different temperatures $T$.
The fit parameters stem from SD and US simulation results, as described in Sec.~\ref{sec:Parametrization in terms of a Gay-Berne potential}, for ${300\,\mathrm{K}}$, ${800\,\mathrm{K}}$, and ${1500\,\mathrm{K}}$.
For temperatures in between we interpolated potential minima and contact distances to receive fit parameters.

\begingroup
\squeezetable
\begin{table}[H]
\begin{ruledtabular}
\caption{\label{tab:M}Model M}
\begin{tabular}{cccccccccccc}
$\mathbf{T}$ $[\mathrm{K}]$&$\boldsymbol\kappa$&$\boldsymbol\chi'$&$\boldsymbol\mu$&$\boldsymbol\nu$&$\boldsymbol\sigma_0$ $[\mathrm{nm}]$&$\boldsymbol\epsilon_0$ $[\mathrm{kJ}/\mathrm{mol}]$&$\mathbf{d}_w$&$\boldsymbol\gamma$&$\boldsymbol\gamma'$&$\boldsymbol\theta$&$\boldsymbol\xi$\\
\hline
&&&&&&&&&&&\\
300&0.2885&-0.7895&1&1&1.0529&6.5481&0.3884&4&4&0.1592&-0.1967\\
400&0.2892&-0.7916&1&1&1.0603&6.3553&0.3869&4&4&0.2019&-0.1984\\
500&0.2899&-0.7939&1&1&1.0678&6.1614&0.3854&4&4&0.2478&-0.2002\\
600&0.2905&-0.7962&1&1&1.0752&5.9666&0.3839&4&4&0.2971&-0.2022\\
700&0.2912&-0.7986&1&1&1.0826&5.7707&0.3825&4&4&0.3503&-0.2043\\
800&0.2919&-0.8012&1&1&1.0900&5.5739&0.3811&4&4&0.4078&-0.2065\\
900&0.2916&-0.8024&1&1&1.0906&5.4425&0.3901&4&4&0.4426&-0.2063\\
1000&0.2914&-0.8037&1&1&1.0912&5.3113&0.3992&4&4&0.4790&-0.2060\\
1100&0.2911&-0.8051&1&1&1.0918&5.1802&0.4082&4&4&0.5171&-0.2057\\
1200&0.2909&-0.8064&1&1&1.0925&5.0494&0.4172&4&4&0.5571&-0.2055\\
1300&0.2907&-0.8079&1&1&1.0931&4.9187&0.4262&4&4&0.5991&-0.2052\\
1400&0.2904&-0.8094&1&1&1.0937&4.7882&0.4352&4&4&0.6434&-0.2049\\
1500&0.2902&-0.8109&1&1&1.0943&4.6579&0.4442&4&4&0.6900&-0.2046\\
\end{tabular}
\end{ruledtabular}
\end{table}
\endgroup
\end{widetext}

\begingroup
\squeezetable
\begin{table}[H]
\begin{ruledtabular}
\caption{\label{tab:K1}Model K1}
\begin{tabular}{cccccccc}
$\mathbf{T}$ $[\mathrm{K}]$&$\boldsymbol\kappa$&$\boldsymbol\kappa'$&$\boldsymbol\mu$&$\boldsymbol\nu$&$\boldsymbol\sigma_0$ $[\mathrm{nm}]$&$\boldsymbol\epsilon_0$ $[\mathrm{kJ}/\mathrm{mol}]$&$\mathbf{d}_w$\\
\hline
&&&&&&&\\
300&0.2885&0.1554&1&1&1.0529&7.7633&0.3884\\
400&0.2892&0.1462&1&1&1.0603&7.1782&0.3869\\
500&0.2899&0.1372&1&1&1.0678&6.6170&0.3854\\
600&0.2905&0.1284&1&1&1.0752&6.0790&0.3839\\
700&0.2912&0.1197&1&1&1.0826&5.5635&0.3825\\
800&0.2919&0.1112&1&1&1.0900&5.0699&0.3811\\
900&0.2916&0.1116&1&1&1.0906&5.0081&0.3901\\
1000&0.2914&0.1121&1&1&1.0912&4.9462&0.3992\\
1100&0.2911&0.1125&1&1&1.0918&4.8843&0.4082\\
1200&0.2909&0.1130&1&1&1.0925&4.8225&0.4172\\
1300&0.2907&0.1135&1&1&1.0931&4.7608&0.4262\\
1400&0.2904&0.1140&1&1&1.0937&4.6990&0.4352\\
1500&0.2902&0.1145&1&1&1.0943&4.6373&0.4442\\
\end{tabular}
\end{ruledtabular}
\end{table}
\endgroup

\begingroup
\squeezetable
\begin{table}[H]
\begin{ruledtabular}
\caption{\label{tab:K2}Model K2}
\begin{tabular}{cccccccc}
$\mathbf{T}$ $[\mathrm{K}]$&$\boldsymbol\kappa$&$\boldsymbol\kappa'$&$\boldsymbol\mu$&$\boldsymbol\nu$&$\boldsymbol\sigma_0$ $[\mathrm{nm}]$&$\boldsymbol\epsilon_0$ $[\mathrm{kJ}/\mathrm{mol}]$&$\mathbf{d}_w$\\
\hline
&&&&&&&\\
300&0.2885&0.0530&1&1&1.0529&2.6836&0.3884\\
400&0.2892&0.0518&1&1&1.0603&2.5722&0.3869\\
500&0.2899&0.0504&1&1&1.0678&2.4602&0.3854\\
600&0.2905&0.0491&1&1&1.0752&2.3476&0.3839\\
700&0.2912&0.0476&1&1&1.0826&2.2345&0.3825\\
800&0.2919&0.0461&1&1&1.0900&2.1209&0.3811\\
900&0.2916&0.0459&1&1&1.0906&2.0748&0.3901\\
1000&0.2914&0.0456&1&1&1.0912&2.0288&0.3992\\
1100&0.2911&0.0453&1&1&1.0918&1.9828&0.4082\\
1200&0.2909&0.0450&1&1&1.0925&1.9369&0.4172\\
1300&0.2907&0.0447&1&1&1.0931&1.8911&0.4262\\
1400&0.2904&0.0444&1&1&1.0937&1.8453&0.4352\\
1500&0.2902&0.0440&1&1&1.0943&1.7996&0.4442\\
\end{tabular}
\end{ruledtabular}
\end{table}
\endgroup
%

\begin{thebibliography}{61}
\bibitem [{bug()}]{bug}%
  \BibitemOpen
  \href@noop {} {}\bibinfo {note} {It was used due to a fatal bug in the
  previous built}\BibitemShut {NoStop}%
\end{thebibliography}
%

\end{document}